%% file: c2-main.tex
\documentclass[fleqn,usenatbib]{mnras}

\usepackage{newtxtext,newtxmath}

\usepackage{multirow}
\usepackage[T1]{fontenc}

\DeclareRobustCommand{\VAN}[3]{#2}
\let\VANthebibliography\thebibliography
\def\thebibliography{\DeclareRobustCommand{\VAN}[3]{##3}\VANthebibliography}

\usepackage{graphicx}	
\usepackage{amsmath}	

\newcommand{\Ha}{H$\alpha$}
\newcommand{\Hb}{H$\beta$}
\newcommand{\HeII}{[He~\textsc{ii}]}
\newcommand{\OII}{[O~\textsc{ii}]}
\newcommand{\CII}{[C~\textsc{ii}]}
\newcommand{\SII}{[S~\textsc{ii}]}
\newcommand{\NII}{[N~\textsc{ii}]}
\newcommand{\OIII}{[O~\textsc{iii}]}
\newcommand{\kms}{\mbox{km s$^{-1}$}}

\newcommand{\sfrsd}{$\Sigma_{\rm SFR}$}

\newcommand{\target}{\mbox{CRISTAL-02}} 
\newcommand{\msun}{M$_\odot$}
\newcommand{\msunyr}{\mbox{M$_\odot$ yr$^{-1}$}}
\newcommand{\msunyrkpc}{\mbox{M$_\odot$ yr$^{-1}$ kpc$^{-2}$}}

\title[A supernova-driven wind in the early Universe]{Multiphase images of a powerful supernova-driven wind in the early Universe}

\author[R. L. Davies et al.]{Rebecca L. Davies,$^{1,2}$\thanks{E-mail: rdavies@swin.edu.au}
D. B. Fisher,$^{1,2}$ R. Herrera-Camus,$^{3,4}$ A. Faisst,$^5$ J. Spilker,$^6$ J. Gonz\'alez-L\'opez,$^{7,8}$ \newauthor
S. Fujimoto,$^{9,\dagger}$ R. Amor\'in,$^{10}$ M. Aravena,$^{8,4}$ R. J. Assef,$^8$ L. Barcos-Mu\~noz,$^{11,12}$ M. Boquien,$^{13}$ \newauthor
M. Dessauges-Zavadsky,$^{14}$ A. Ferrara,$^{15}$ N. M. F\"orster Schreiber,$^{16}$ M. Ginolfi,$^{17,18}$ D. G\'{o}mez-Espinoza,$^{19}$ \newauthor
E. Ibar,$^{19,4}$ R. Ikeda,$^{20,21}$ H. Inami,$^{22}$ G. C. Jones,$^{23,24}$ A. Koekemoer,$^{25}$ L. L. Lee,$^{16}$ J. Li,$^{26}$ D. Liu,$^{27}$ \newauthor
Z. Liu,$^{28,29,30,31}$ I. De Looze,$^{32}$ I. Mitsuhashi,$^{33}$ J. Molina,$^{19}$ A. Nanni,$^{34,35}$ M. Relano,$^{36,37}$ \newauthor
M. Romano,$^{38,39}$ P. Sawant,$^{34}$ M. Solimano,$^8$ L. Sommovigo,$^{40}$ K. Tadaki,$^{41}$ K. Telikova,$^8$ H. \"Ubler,$^{16}$ \newauthor
V. Villanueva,$^3$ W. Wang,$^5$ and G. Zamorani$^{42}$.
\\
$^1$Centre for Astrophysics and Supercomputing Swinburne University of Technology, John Street Hawthorn, 3122, VIC, Australia \\
$^2$ARC Centre of Excellence for All Sky Astrophysics in 3 Dimensions (ASTRO 3D) \\
$^3$Departamento de Astronom\'ia, Universidad de Concepci\'on, Barrio Universitario, Concepci\'on, Chile \\
$^4$Millenium Nucleus for Galaxies \\
$^5$Caltech/IPAC, 1200 E. California Blvd. Pasadena, CA 91125, USA \\
$^6$Department of Physics and Astronomy, Texas A\&M University, 4242 TAMU, College Station, TX 77843-4242, USA \\
$^7$Las Campanas Observatory, Carnegie Institution of Washington, Casilla 601, La Serena, Chile \\
$^8$Instituto de Estudios Astrof\'isicos, Facultad de Ingenier\'ia y Ciencias, Universidad Diego Portales, Av. Ej\'ercito Libertador 441, Santiago, Chile \\
$^9$Department of Astronomy, The University of Texas at Austin, Austin, TX 78712, USA \\
$^{10}$Instituto de Astrof\'{i}sica de Andaluc\'{i}a (CSIC), Apartado 3004, 18080 Granada, Spain \\
$^{11}$National Radio Astronomy Observatory, 520 Edgemont Road, Charlottesville, VA 22903, USA \\
$^{12}$Department of Astronomy, University of Virginia, 530 McCormick Road, Charlottesville, VA 22903, USA \\
$^{13}$Université Côte d'Azur, Laboratoire Lagrange, Nice, France \\
$^{14}$Department of Astronomy, University of Geneva, Chemin Pegasi 51, 1290 Versoix, Switzerland \\
$^{15}$Scuola Normale Superiore, Piazza dei Cavalieri 7, 50126 Pisa, Italy \\
$^{16}$Max-Planck-Institut für extraterrestrische Physik, Gießenbachstraße 1, D-85748 Garching, Germany \\
$^{17}$Universit\`a di Firenze, Dipartimento di Fisica e Astronomia, via G. Sansone 1, 50019 Sesto Fiorentino, Florence, Italy \\
$^{18}$INAF -- Arcetri Astrophysical Observatory, Largo E. Fermi 5, I-50125, Florence, Italy \\
$^{19}$Instituto de F\'{i}sica y Astronom\'{i}a, Universidad de Valpara\'{i}so, Avda. Gran Breta\~{n}a 1111, Valpara\'{i}so, Chile \\
$^{20}$Department of Astronomy, School of Science, SOKENDAI (The Graduate University for Advanced Studies), 2-21-1 Osawa, Mitaka, Tokyo 181-8588, Japan \\
$^{21}$National Astronomical Observatory of Japan, 2-21-1 Osawa, Mitaka, Tokyo 181-8588, Japan \\
$^{22}$Hiroshima Astrophysical Science Center, Hiroshima University, 1-3-1 Kagamiyama, Higashi-Hiroshima, Hiroshima 739-8526, Japan \\
$^{23}$Kavli Institute for Cosmology, University of Cambridge, Madingley Road, Cambridge CB3 0HA, UK \\
$^{24}$Cavendish Laboratory, University of Cambridge, 19 JJ Thomson Avenue, Cambridge CB3 0HE, UK \\
$^{25}$Space Telescope Science Institute, 3700 San Martin Drive, Baltimore, MD 21218, USA \\
$^{26}$International Centre for Radio Astronomy Research (ICRAR), The University of Western Australia, M468, 35 Stirling Highway, Crawley, WA 6009, Australia \\
$^{27}$Purple Mountain Observatory, Chinese Academy of Sciences, 10 Yuanhua Road, Nanjing 210023, China \\
$^{28}$Kavli Institute for the Physics and Mathematics of the Universe (WPI), UTIAS, Tokyo Institutes for Advanced Study, University of Tokyo, Chiba, 277-8583, Japan \\
$^{29}$Department of Astronomy, School of Science, The University of Tokyo, 7-3-1 Hongo, Bunkyo, Tokyo 113-0033, Japan \\
$^{30}$Center for Data-Driven Discovery, Kavli IPMU (WPI), UTIAS, The University of Tokyo, Kashiwa, Chiba 277-8583, Japan \\
$^{31}$Université Paris-Saclay, Université Paris Cité, CEA, CNRS, AIM, F-91191 Gif-sur-Yvette, France \\
$^{32}$Sterrenkundig Observatorium, Ghent University, Krijgslaan 281 - S9, B-9000 Gent, Belgium \\
$^{33}$Department for Astrophysical \& Planetary Science, University of Colorado, Boulder, CO 80309, USA \\
$^{34}$National Centre for Nuclear Research, ul. Pasteura 7, 02-093 Warsaw, Poland \\
$^{35}$INAF - Osservatorio astronomico d'Abruzzo, Via Maggini SNC, 64100, Teramo, Italy \\
$^{36}$Dept. Física Te\'{o}rica y del Cosmos, Campus de Fuentenueva, Edificio Mecenas, Universidad de Granada, E-18071, Granada, Spain \\
$^{37}$Instituto Universitario Carlos I de Física Te\'{o}rica y Computacional, Universidad de Granada, 18071, Granada, Spain \\
$^{38}$Max-Planck-Institut für Radioastronomie, Auf dem Hügel 69, 53121 Bonn, Germany \\
$^{39}$INAF - Osservatorio Astronomico di Padova, Vicolo dell’Osservatorio 5, I-35122 Padova, Italy \\
$^{40}$Center for Computational Astrophysics, Flatiron Institute, 162 5th Avenue, New York, NY 10010, USA \\
$^{41}$Faculty of Engineering, Hokkai-Gakuen University, Toyohira-ku, Sapporo 062-8605, Japan \\
$^{42}$INAF – Osservatorio di Astrofisica e Scienza dello Spazio di Bologna, Via Gobetti 93/3, 40129 Bologna, Italy \\
$^\dagger$Hubble Fellow
}

\date{Accepted XXX. Received YYY; in original form ZZZ}

\pubyear{\the\year{}}

\begin{document}
\label{firstpage}
\pagerange{\pageref{firstpage}--\pageref{lastpage}}
\maketitle

\begin{abstract}
Galactic winds are considered a likely driver of rapid quenching in early massive galaxies, but until now there has been no direct evidence that such systems drive winds powerful enough to meaningfully suppress their star-formation. We present resolved cold gas and ionized gas observations of a powerful supernova-driven wind in a massive galaxy 1.1 billion years after the Big Bang (at $z$=5.3). The outflow, likely triggered by ongoing merger activity, is removing gas at twice the rate of star-formation and could plausibly eject all the cold gas from the galaxy within 100~Myr. Our results suggest that powerful merger-driven outflows may be a key mechanism to produce abundant massive quiescent galaxies in the early Universe when a large fraction of massive galaxies are interacting. The mass and energetics of this distant outflow are consistent with nearby starburst-driven superwinds, suggesting that the efficiency of stellar feedback has remained relatively constant over the last 12 billion years of cosmic history.
\end{abstract}




\section{Introduction}

One of the most surprising discoveries from the James Webb Space Telescope (JWST) is that massive quiescent galaxies were already present in large numbers at $3 \lesssim z\lesssim$~5, when the Universe was only \mbox{1~--~2~billion} years old \citep[e.g.][]{Glazebrook24, DeGraaff25}. Observations and simulations suggest that these galaxies built most of their mass in strong bursts of star-formation that were immediately followed by rapid quenching \citep[e.g.][]{Nanayakkara25}. However, the properties of their progenitors and the mechanisms that trigger the starbursts and subsequent quenching remain poorly characterised in observations. Powerful galaxy-scale outflows can deplete their host galaxies of fuel for star-formation within a few hundred Myr and may thus play an important role in forming the first massive quiescent galaxies \citep[e.g.][]{Spilker18, Belli24}. To evaluate the likelihood of this scenario, it is essential to determine whether sufficiently powerful winds are present in typical massive galaxies within the first billion years of the Universe.

\begin{sloppypar}
Observations of the \CII~$\lambda$158$\mu$m emission line have provided key insights into the properties of early outflows. With an ionization potential of 11.3~eV, \CII\ can trace ionized, atomic or molecular gas. In the dense interstellar medium (ISM), \CII\ primarily arises from photodissociation regions at the outer layers of molecular clouds \citep[e.g.][]{Hollenbach99}, whilst in more diffuse outflows and extraplanar gas, \CII\ mainly traces atomic gas \citep{Madden93, Levy23}. Stacking analyses of typical main sequence galaxies at $z\sim$~5 have revealed evidence for weak broad \CII\ emission components that may trace outflows \citep[e.g.][]{Gallerani18, Ginolfi20a, Birkin25}. Furthermore, many $z\sim$~5 galaxies show extended \CII\ `halos' whose properties can be explained by catastrophically cooling outflows \citep[e.g.][]{Pizzati20}. However, neither broad emission line components or spatially extended \CII\ emission are unambiguous signatures of outflowing gas: both of these could plausibly be explained by tidal tails or stripped material in interacting systems \citep[e.g.][]{Ginolfi20b}. 
\end{sloppypar}

Optical emission lines provide valuable constraints on the nature of gas flows in interacting systems. In the local Universe, almost all starburst-driven superwinds are in interacting galaxies, and there are well-established methods to distinguish between outflows and other interaction-induced flows using the kinematics, emission line ratios and spatial profiles of optical emission lines \citep[e.g.][]{Baron24}. With JWST, it is now possible to observe rest-frame optical emission lines at $z\sim$~5, providing a means to robustly identify outflows and examine their impact on star formation in early massive galaxies.

\begin{figure*}
    \centering
    \includegraphics[scale = 0.55]{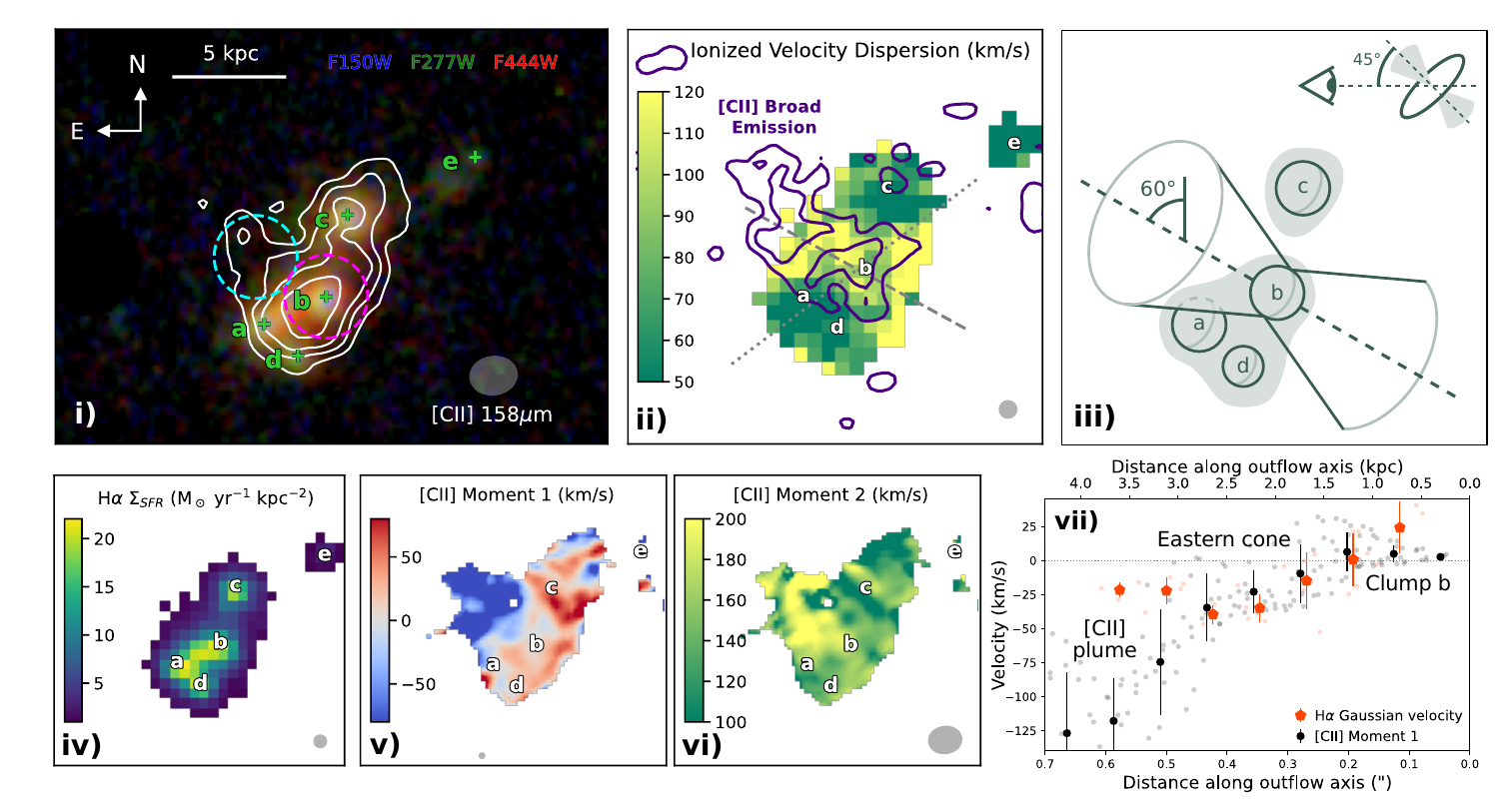} \\
    \caption{\textbf{JWST and ALMA observations of \target}. 
    i) 4~$\times$~3'' JWST/NIRCam 3-color composite image with the star-forming clumps (a-d) and faint companion (e) marked (`+` signs indicate clump centres). Contours show the \CII~$\lambda$158$\mu$m flux at levels of 4, 7, 10, 15$\sigma$, and reveal a plume to the NE of Clump~b. The dashed magenta and cyan circles show the apertures used to extract the spectra of Clump b and the plume, respectively. Grey ellipses in the bottom right of panels i)~--~vi) show the point spread function of the observations. ii) Ionized gas velocity dispersion (measured from 1-component fitting and corrected for instrumental dispersion) with contours showing the broad, blueshifted \CII\ emission (integrated over $-$500 to $-$150~\kms) at levels of 3 and 5$\sigma$. The broad emission is enhanced in a biconical structure along the NE-SW axis, characteristic of outflowing gas. The dotted line indicates the kinematic major axis (PA = 127$^\circ$), the dashed line delineates the outflow axis \mbox{(PA = 60$^\circ$)}. iii) Cartoon showing the geometry of the outflow, which likely originates from Clump b. iv)~--~vi) Maps of star formation rate surface density from \Ha, and \CII\ moment 1 (velocity offset) and moment 2 (velocity width). vii) Velocity profiles of \CII\ (black) and \Ha\ (orange) emission along the outflow axis. The gas becomes increasingly blueshifted moving from Clump b towards the \CII\ plume. }
    \label{fig:gallery}
\end{figure*}

\section{Data and Results}
We present resolved \CII\ and rest-frame optical emission line observations of outflowing gas in \target, a galaxy system with stellar mass $M_* \simeq$~2~$\times$~10$^{10}$~\msun\ \citep{Mitsuhashi24}, at a redshift \mbox{$z_{\rm CII} = 5.294$}. Based on the \Ha\ luminosity, \target\ forms stars at a rate of 260~$\pm$~30~\msunyr, about three times the average value for `main sequence' star-forming galaxies at similar mass and redshift \citep[e.g.][]{Speagle14}. Resolved at $\sim$~400~pc scale by JWST, the system is dominated by several massive star-forming clumps (Fig.~\ref{fig:gallery}~i,~iv). The three southern clumps (a, b and d) form a coherent kinematic structure that roughly resembles a rotating disk with a position angle (PA) of 127$^\circ$, whilst the northern clump (c) and the faint companion (e) are kinematically distinct (see Supplementary Material), suggesting \target\ is likely an early-stage merger (see also~\citealt{Capak15, Guaita22}). 

ALMA \CII\ observations were obtained as part of the CRISTAL \citep{HerreraCamus25} and ALPINE \citep{LeFevre20} surveys as well as programs 2012.1.00523.S \citep{Capak15} and 2011.0.00064.S \citep{Riechers14}. A continuum-subtracted datacube was produced using Briggs weighting with 20~\kms\ velocity resolution, resulting in a final beam full width at half maximum of 0.29~$\times$~0.36'' (1.8~$\times$~2.2 kpc at $z\sim$~5.3) and an rms of 160 $\mu$Jy beam$^{-1}$. Rest-frame optical emission lines were observed with the JWST/NIRSpec integral field spectrograph as part of program GO-3045 \citep{Faisst26}, covering \Ha, \Hb, \NII~$\lambda$6583\AA, and \OIII~$\lambda$5007\AA\ with a spectral resolution of $R = \lambda/\Delta \lambda \simeq$~1000 ($\Delta v \simeq$~300~km~s$^{-1}$). 

\begin{figure*}
    \centering
    \includegraphics[scale = 0.65, clip = True, trim = 15 65 18 0]{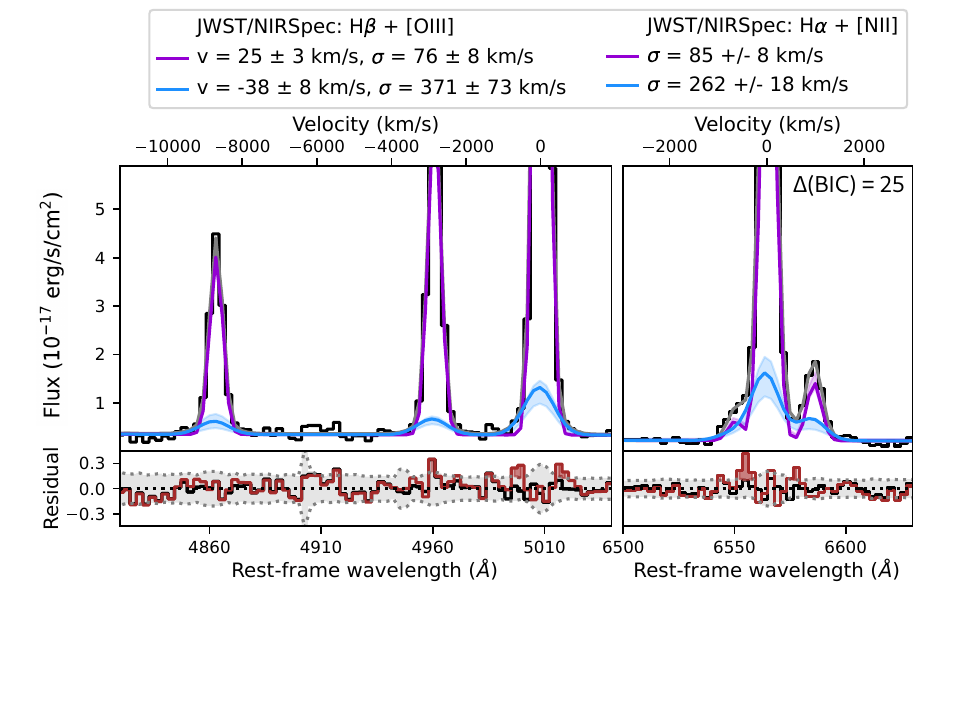} \includegraphics[scale = 0.65, clip = True, trim = 0 65 225 0]{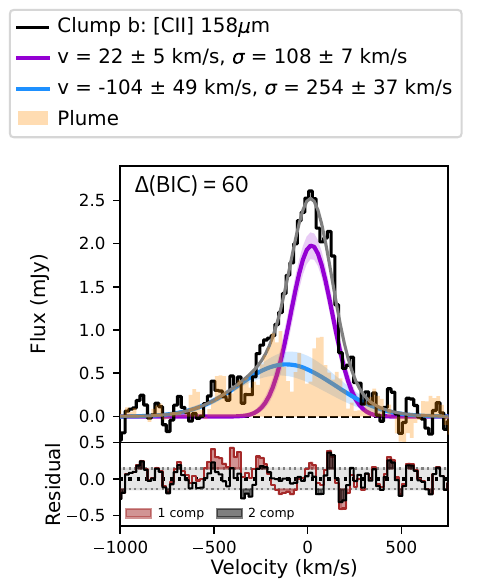}    
    \caption{\textbf{Spectra of the outflow}, extracted in a 0.6'' (3.7 kpc) diameter aperture centered on Clump~b. The emission line profiles are well fit by two Gaussian components: a narrow ISM component (purple) and a broader, blueshifted outflow component (blue). The blue and purple shaded regions represent the 16th-84th percentile ranges from the MCMC posteriors and the grey curves show the total fits. In the lower panels, the grey shading shows the 1$\sigma$ error region. Black lines show the residuals of the two-component fits and black shading highlights residuals above 1$\sigma$, whilst brown lines and shading indicate the larger residuals if only a single Gaussian component is fit. The reported velocity dispersion values are after removing the instrumental line broadening. In the right panel, the filled orange region shows the spectrum extracted from the plume, which is very similar in shape and amplitude to the fitted broad component of Clump~b.}
    \label{fig:multicomp_fits}
\end{figure*}

We detect a plume of \CII\ emission that extends 7~kiloparsecs (kpc) to the north-east of \target\ (Fig.~\ref{fig:gallery}~i). The gas in this plume is blueshifted by $\sim$100~\kms\ (Fig.~\ref{fig:gallery}~v) and has a large velocity width (second moment $M_{2,v} \simeq$~200~\kms) that remains approximately constant with distance (Fig.~\ref{fig:gallery}~vi), consistent with predictions for an outflow in which both the intrinsic turbulence and the volume of the outflow cone probed along the line of sight should increase with distance \citep[e.g.][]{Schneider20}. The broad, blueshifted, \CII-emitting gas is centered on the highly star-forming Clump~b and oriented in a quasi-biconical structure almost perpendicular to the kinematic major axis of \target; a key characteristic of starburst-driven outflows in the local Universe \citep[e.g.][]{Bik18, McPherson23}. The plume is not seen in the shallower ionized gas maps, but the ionized and neutral gas share many important attributes. The ionized gas velocity dispersion is enhanced in a striking biconical feature that aligns closely with the broad \CII\ emission (Fig.~\ref{fig:gallery}~ii) and shows elevated \NII/\Ha\ ratios indicative of shock excitation (see Supplementary Material); similar to nearby starburst-driven winds \citep[e.g.][]{Sharp10}. The \CII\ and \Ha\ line profiles in Clump~b both show broad, blueshifted wings with similar velocity widths (Fig. \ref{fig:multicomp_fits}), and both phases trace the same negative velocity gradient between Clump~b and the eastern edge of the bicone (Fig.~\ref{fig:gallery}~vii). These similarities strongly suggest that \Ha\ and \CII\ probe the same multiphase gas (Fig.~\ref{fig:gallery}~iii).

An outflow is the most natural explanation for blueshifted, shock-excited gas that originates from a highly star-forming clump and propagates along the minor axis whilst maintaining high velocity dispersion to large radii. Tidal tails can produce extraplanar \CII\ emission but typically originate from the outer regions of galaxies and often show decreasing velocity dispersion profiles (see e.g.~\citealt{Duc13}). Furthermore, our subsequent analysis shows that the mass and velocity of the \CII\ plume are consistent with expectations for a supernova-driven outflow. A tidal tail would require very fine-tuned mass, kinematics and geometry to reproduce the \CII\ plume properties, making this scenario very unlikely. The absence of strong \CII\ emission from the western side of the outflow may indicate that the outflow is asymmetric (perhaps due to the ongoing galaxy-galaxy interactions; e.g.~\citealt{McPherson23}), or the receding side of the outflow may be observed in projection against the host galaxy \citep[e.g.][]{Bik18}. The detection of spatially resolved, outflowing gas in \CII\ and rest-frame optical emission lines gives us a unique multiphase view of ejective feedback in this distant galaxy. 

Motivated by the distinctive biconical feature, we define outflowing pixels as those exceeding a given ionized gas velocity dispersion threshold. We test thresholds of $\sigma >$ 85, 90, 95, and 100~\kms\ and estimate the opening angle assuming that the outflow consists of two truncated cones, following \citet{McPherson23}. We obtain full opening angles of 30~--~55$^\circ$ for the eastern cone and \mbox{55~--~65$^\circ$} for the western cone. The fact that we observe both a biconical structure and broad line emission (Fig.~\ref{fig:multicomp_fits}) suggests that the outflow is observed at intermediate inclination ($i \simeq$~30~--~60~deg; e.g.~\citealt{Westmoquette11}); consistent with the galaxy inclination measured from the JWST/NIRCam image ($i$~=~47$^\circ$; see Supplementary Material).

The most likely source of the outflow is Clump b which lies at the centre of the bicone. Most of the broad \CII\ emission originates from Clump~b, the plume, and the area between them (Fig.~\ref{fig:gallery}~ii), and the amplitude and velocity profile of the emission are consistent across this region (Fig.~\ref{fig:multicomp_fits}). Clump~b has a peak \Ha\ star-formation rate surface density (\sfrsd) of \mbox{21~\msunyrkpc} (Fig.~\ref{fig:gallery}~iv); easily enough to power an outflow \citep[e.g.][]{Heckman02, Davies19}. To quantify the outflow properties, we extract \CII\ and rest-frame optical spectra over a 0.6'' (3.7~kpc) diameter aperture centered on Clump~b (magenta dashed circle in Fig.~\ref{fig:gallery}~i); covering roughly the width of the bicone. We fit each emission line with two Gaussian components: a narrow ISM component and a broad outflow component. For the rest-frame optical emission lines, we also tried fitting a broad component solely to \Ha, but found that the Bayesian Information Criterion (BIC) supports adding broad components to the \NII\ lines as well \mbox{($\Delta$(BIC) = 10 relative to the broad \Ha\ model)}.

We measure the maximum projected outflow velocity using the kinematics of the outflow component \mbox{($v_{\rm out,max}$ = $|\Delta v|$ + 2$\sigma$)}, yielding \mbox{$v_{\rm out,max}$ = 640$^{+70}_{-60}$ \kms} for \CII, \mbox{550$^{+80}_{-70}$~\kms} for \Ha\ and \mbox{740$^{+150}_{-120}$~\kms} for \OIII. The \CII-emitting neutral gas and optically-emitting ionized gas have similar velocities, suggesting that the two phases are kinematically coupled. The \CII\ spectra have substantially finer velocity resolution than the optical spectra (20~\kms\ compared to $\sim$300~\kms; see Fig.~\ref{fig:multicomp_fits}), so we adopt the \CII\ outflow velocity as the fiducial value in our subsequent analysis. The escape velocity from \target\ is $\sim$600~\kms\ (see Supplementary Material) which is comparable to the (line of sight) outflow velocity, suggesting that a non-negligible fraction of the outflowing gas could leave the system. Accounting for the likely outflow inclination of 30~--~60~deg yields deprojected outflow velocities of \mbox{740~--~1280~\kms}. In contrast, tidal tails typically have velocities comparable to the rotation velocity at the outer edge of the galaxy.

We compute the total outflowing mass traced by the \CII\ broad emission component assuming that this gas is primarily neutral, following \citet{HerreraCamus21}. Most \CII\ emission from outflows is expected to trace neutral gas \citep[e.g.][]{Madden93, Levy23}, and the \CII/\NII$\lambda$205$\mu$m ratio in \target\ implies that $>$~90\% of the \CII-emitting material is neutral \citep{Pavesi19}. Adopting the most conservative \CII\ mass-to-light ratio, we obtain an outflow mass of \mbox{$M_{\rm out, [CII]}$ = 1.5~$\pm$~0.3 $\times$ 10$^9$ \msun} (see Supplementary Material). We then estimate the ionized outflow mass using the extinction-corrected broad \Ha\ luminosity. We adopt an electron density of $n_e$~=~100~cm$^{-3}$, similar to both the density measured from the \SII\ doublet ratio in clump~b (220~cm$^{-3}$) and the density of the M82 outflow \citep{XuXinFeng23_M82}. The derived mass is \mbox{$M_{\rm out, ion} =$ 2.5$^{+0.9}_{-0.7}~ \times$ 10$^8$ (100~cm$^{-3}$/$n_e$) \msun}. The ionized phase is subdominant for all realistic electron density estimates, with a mass fraction of \mbox{15\% $\times$ (100~cm$^{-3}$/$n_e$)}. Assuming a mass-conserving biconical outflow, and adopting an outflow size of 1.9~kpc (the radius of the extraction aperture), the total outflow rate of \CII-emitting gas is \mbox{521$^{+71}_{-89}$~\msunyr} and the implied mass-loading factor \mbox{($\eta$ = $\dot{M}_{\rm out}$/SFR)} is 2.0~$\pm$~0.4; consistent with predictions from chemical evolution modelling of massive $z\sim$~5 galaxies \citep{Sawant25}.

\begin{figure*}
    \centering
    \includegraphics[scale = 0.82, clip = True, trim = 5 140 10 0]{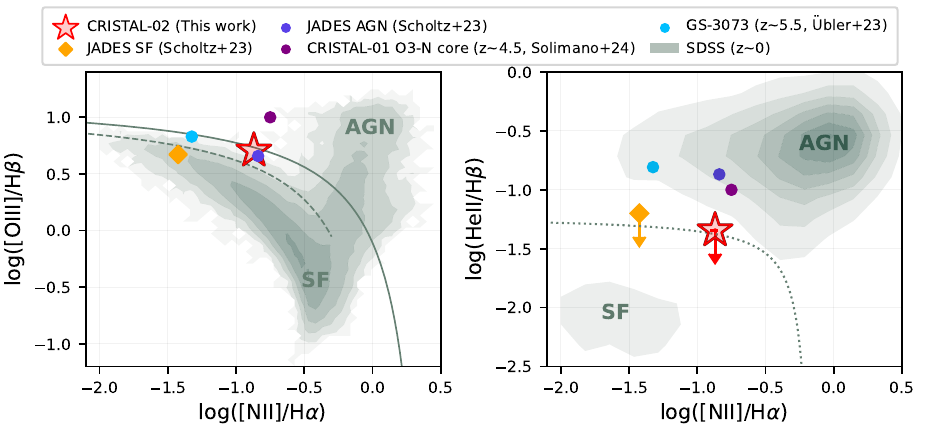} 
    \caption{\textbf{Emission line ratios of \target\ (red star)}, compared with $z\sim$~0 galaxies (filled contours, \citealt{Shirazi12}) and $z\sim$~3~--~6 star-forming galaxies (orange diamond) and AGN hosts (blue/purple circles). Left: \NII/\Ha\ vs. \OIII/\Hb\ diagram. The dashed curve is the locus of $z\sim$~2 star-forming galaxies \citep{Strom17} and the solid curve shows the maximum line ratios attributable to pure star-formation \citep{Ke01a}. Right: \NII/\Ha\ vs. \HeII/\Hb\ diagram. The dotted curve shows the demarcation between star-formation and AGN dominated sources \citep{Shirazi12}. Low metallicity AGN occupy the same region as star-forming galaxies in the \NII/\Ha\ vs. \OIII/\Hb\ diagram, but the two populations are clearly separated in \HeII/\Hb. We measure a sensitive upper limit on the \HeII\ flux from \target, indicating that AGN activity is unlikely to be energetically significant.}
    \label{fig:bpt}
\end{figure*}

Finally, we estimate the molecular gas mass of \target\ using the narrow (galaxy) \CII\ luminosity. The ISM of distant galaxies is primarily molecular, and as a result, \CII\ emission from the ISM is a widely used tracer of molecular gas mass \citep[e.g.][]{Zanella18, DessaugesZavadsky20, Casavecchia25}. We extract an integrated \CII\ spectrum covering the entire region where \CII\ is detected at $\geq$~4$\sigma$ significance (see Fig.~\ref{fig:gallery}~i) and fit two Gaussian components to separate the narrow emission of the host galaxy from the broad emission of the outflow. We estimate that \target\ has a molecular gas mass of \mbox{$\log(M_{\rm mol}/$\msun)~=~10.3~--~10.6}; consistent with estimates based on the Band 7 continuum (see Supplementary Material) and the non-detection of CO(2-1) emission \citep{Pavesi19}. This implies a molecular-to-baryonic mass fraction of 0.5~--~0.67; similar to typical main sequence galaxies at this redshift \citep[e.g.][]{DessaugesZavadsky20}. Assuming that the outflow continues at the same rate and there is no gas replenishment from the halo, the outflow could eject the entire molecular gas reservoir of the host galaxy in less than 100~Myr; about twice as fast as the depletion timescale for conversion of this gas into stars.

Notably, the powerful outflow from \target\ seems most likely to be powered by star-formation. \target\ is not detected in X-ray or radio continuum, the multiwavelength photometry is consistent with pure star-formation (\citealt{Li24}; see Supplementary Material), and the kinematic profiles of the Balmer lines in Clump~b do not show broadening characteristic of a Type~1 active galactic nucleus (AGN; Fig.~\ref{fig:multicomp_fits}). Furthermore, the \NII/\Ha, \OIII/\Hb\ and \HeII/\Hb\ emission line ratios are consistent with ionization by young stars (Fig.~\ref{fig:bpt}). The rate of energy injection by supernova explosions is comparable to the outflow kinetic power (\mbox{$\dot{E}_{\rm SN}$~$\simeq$~$\dot{E}_{\rm out} \simeq$~6~--~7~$\times$~10$^{43}$~erg s$^{-1}$}; see Supplementary Material), consistent with the energetics measured for nearby starburst-driven superwinds \citep[e.g.][]{Thompson24}. Finally, the high mass-loading factor ($\eta \simeq$~2) is consistent with predictions for catastrophically cooling starburst-driven winds \citep{Pizzati20}. The properties of the \target\ outflow are listed in Table \ref{table:outflow_props}.

\section{Discussion and Conclusions}
Measuring the resolved, multiphase properties of a $z\sim$~5.3 outflow provides a unique opportunity to test whether the physics governing star-formation-driven outflows differs significantly at high redshifts. JWST has revealed an over-abundance of UV-bright galaxies at $z >$~6 and a high number density of massive quiescent galaxies at $z\sim$~3~--~4, suggesting that galaxies grow much more rapidly than expected in the early Universe \citep[e.g.][]{Harikane25b}. This could point to less efficient feedback in extreme high-redshift environments. Any evolution in how feedback operates over cosmic time would have important implications for cosmological simulations. 

\begin{figure*}
    \centering
    \includegraphics[scale = 0.78, clip = True, trim = 10 0 10 0]{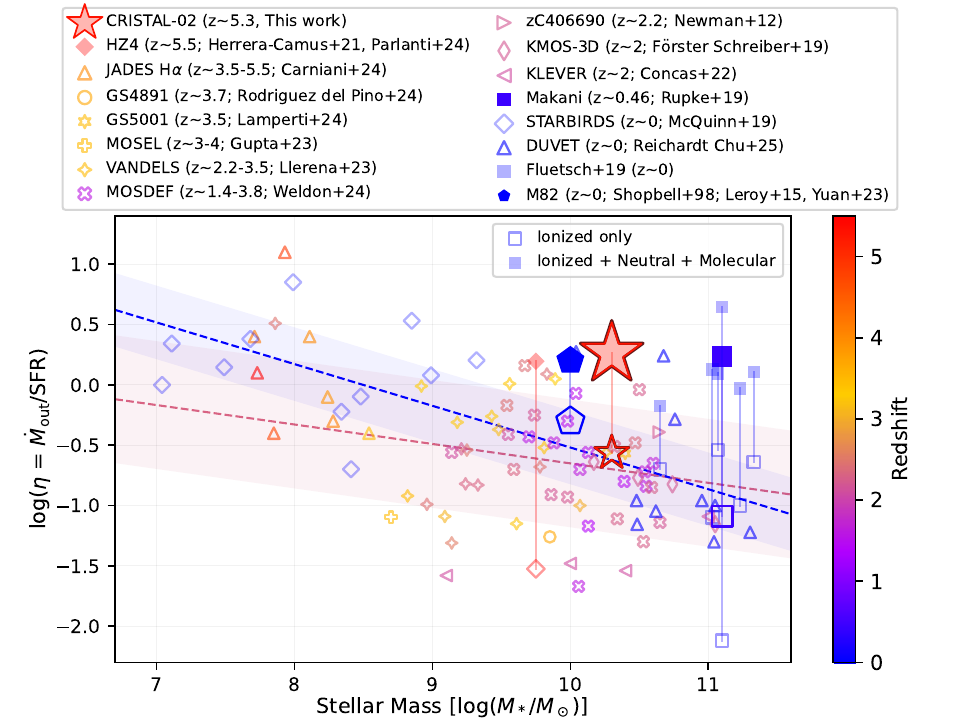} \includegraphics[scale = 0.76, clip = True, trim = 10 0 240 0]{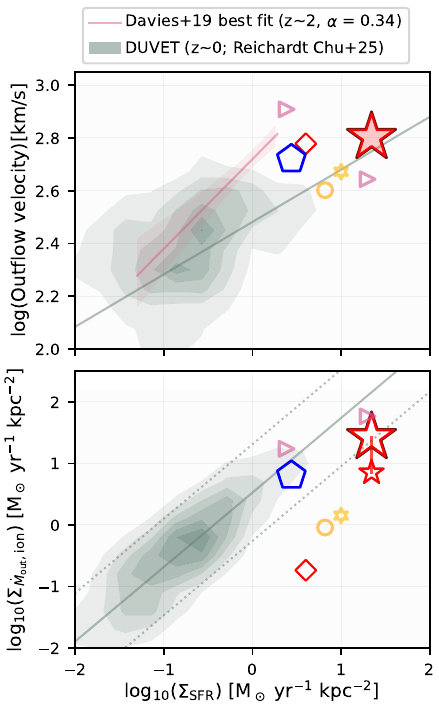} 
    \vspace{-10pt}
    \caption{\textbf{Scaling relations between galaxy properties and outflow properties. Left: Outflow mass loading factor as a function of stellar mass} for 99 star-formation-driven outflows spanning 12 billion years of cosmic time. Literature samples are described in Table \ref{table:literature_sample}. Open symbols show measurements for the ionized phase, and filled markers indicate total outflow rates for 10 galaxies with multiphase measurements. Dashed lines and shaded regions show the best-fit linear scalings for the ionized phase at $z\sim$~0 (blue) and 2~$\lesssim z\lesssim$~5.5 (pink) and their associated 1$\sigma$ errors. \target\ shows similar outflow properties to comparably massive galaxies at lower redshifts and there is no evidence for evolution in outflow mass-loading factor with cosmic time. Right: Outflow velocity (top) and ionized outflow mass flux (bottom) as a function of \sfrsd\ for outflows with spatially resolved measurements. Filled grey contours show measurements of 500~pc regions from $z\sim$~0 starburst galaxies and grey lines show the associated best fits and 2$\sigma$ scatter. We show two outflow mass flux values for \target: the small star is the fiducial Clump~b measurement, and the larger star assumes $R_{\rm out}$~=~500~pc for consistency with the $z\sim$~0 values. \target\ follows similar velocity and mass trends to lower redshift outflows but at much higher \sfrsd, providing a unique test of supernova-driven outflow models. }
    \label{fig:scaling_relations}
\end{figure*}

To test whether the feedback strength evolves with redshift, we collated a sample of 99 star-formation-driven outflows from the literature spanning 12 billion years of cosmic history (see Table \ref{table:literature_sample}). We focus on outflows measured in the ionized phase which is the most commonly detected at both low and high redshifts. 9/99 (9\%) outflows have also been measured in the cold atomic/molecular phase, and 16/99 (16\%) have spatially resolved observations. The left-hand panel of Fig. \ref{fig:scaling_relations} shows the outflow mass loading factor as a function of stellar mass \citep[e.g.][]{Nelson19, Pandya21}, with \target\ marked by the red star. Open symbols show measurements for the ionized phase, and filled markers add the neutral/molecular phase when available. Considering only ionized gas measurements, we measure a slope of \mbox{$\alpha \simeq -0.3$}; slightly flatter than the theoretical prediction of \mbox{$\alpha \simeq -0.5$} \citep[e.g.][]{Pandya21}. The correlation does not evolve significantly with redshift: the best-fit relations for outflows at \mbox{$z < 0.2$} (blue dashed line and shaded region) and \mbox{2.0 $\lesssim z \lesssim$~5.5} (pink) are indistinguishable within the 1$\sigma$ errors. \target\ shows very similar properties to M82 and other massive $z\sim$~0 galaxies.

We examine how the starburst energy couples to the outflowing gas by plotting the outflow velocity (Fig.~\ref{fig:scaling_relations} top right) and ionized outflow mass flux ($\Sigma_{\dot{M}_{\rm out},{\rm ion}}$, bottom right) as a function of the \sfrsd\ at the outflow launch point, for the subset of outflows with spatially resolved measurements. Filled contours and solid lines show measurements and corresponding best fits for 500~pc regions in $z\sim$~0 starburst galaxies. The velocity of the \target\ outflow is consistent with the extrapolation of the $z\sim$~0 trend to higher \sfrsd. The outflow mass flux falls slightly below the $z\sim$~0 relation, but is within the 2$\sigma$ scatter on the measurements. The offset could easily be explained by systematic uncertainties on the outflow electron density, size, and dust extinction, which combined lead to order-of-magnitude uncertainties on both the $z\sim$~0 and \target\ values \citep[e.g.][]{ReichardtChu22}. Our results suggest that the efficiency of star-formation-driven outflows has remained similar over much of cosmic history despite significant evolution in the internal properties of galaxies. 

High resolution outflow simulations predict that the distribution of outflowing mass between gas phases varies significantly with \sfrsd\ \citep[e.g.][]{Rathjen23}. 9/99 outflows in the literature sample have ionized and neutral gas measurements (filled markers in Fig.~\ref{fig:scaling_relations} left), enabling us to compute their ionized mass fractions. Although these mass fractions have significant associated uncertainties, we find that galaxies with similar \sfrsd\ to \target\ (including nearby starbursts and the $z=$~0.46 galaxy `Makani') have ionized outflow mass fractions of \mbox{5~--~30\%} \citep[e.g.][]{Fluetsch19, Rupke19, Yuan23}; comparable to the $\sim$~15\% we measure for \target. In contrast, $z\sim$~0 main sequence star-forming galaxies with 10~--~100 times lower \sfrsd\ also have lower ionized fractions of \mbox{$<$~1~--~6\%} \citep[e.g.][]{RobertsBorsani20,Avery22}. A recent study reported an ionized outflow fraction of 2\% in a $z\sim$~5.5 star-forming galaxy \citep{HerreraCamus21, Parlanti25}, which is unexpected given its high \sfrsd. However, the \CII\ and rest-frame optical emission line profiles in this galaxy have different kinematic profiles and likely do not trace the same gas, complicating the interpretation of the ionized mass fraction in this system. 

\renewcommand{\arraystretch}{1.2} 
\begin{table}
    \centering
    \caption{\textbf{Outflow properties} measured using ALMA [C~\textsc{ii}] data (upper) and JWST/NIRSpec ionized gas data (lower).}
    \begin{tabular}{l|l}
    \textbf{Property} & \textbf{Value} \\ \hline
    \multicolumn{2}{c}{Measurements based on ALMA \CII\ data} \\ \hline
    $v_{\rm out}$ & 640$^{+70}_{-60}$ km/s \\
    $L_{\rm [CII], broad}$ & 1.2$^{+0.3}_{-0.2}$~$\times$~10$^{42}$ erg s$^{-1}$ \\
    $M_{\rm out}$ & (1.5~$\pm$~0.3) $\times$ 10$^9$ \msun\ \\ 
    $\dot{M}_{\rm out}$ & 521$^{+71}_{-89}$ \msunyr\ \\
    $\Sigma_{\dot{M},{\rm out}}$ & 47$^{+7}_{-8}$ \msunyrkpc\ \\
    $\eta$ & 2.0~$\pm$~0.4 \\ 
    $\dot{E}_{\rm out}$	& 6.6~$\pm$~1.4 $\times$ 10$^{43}$ erg s$^{-1}$ \\ 
    $\dot{E}_{\rm SN}$ & 6.2~$\pm$~1.1~$\times$~10$^{43}$ erg s$^{-1}$ \\ \hline
    \multicolumn{2}{c}{Measurements based on JWST/NIRSpec ionized gas data} \\ \hline
    Opening angle - western cone & 55~--~65 deg \\
    Opening angle - eastern cone & 30~--~55 deg \\
    $v_{\rm out}$(\Ha)	& 550$^{+80}_{-70}$ km/s \\
    $v_{\rm out}$(\OIII) & 740$^{+150}_{-120}$ km/s \\
    $L_{H\alpha, \rm broad}$ & 7.8$^{+2.8}_{-2.2}$ $\times$~10$^{42}$ erg s$^{-1}$ \\
    $M_{\rm out}$(ionized) & $2.5^{+0.9}_{-0.7}$ $\times$ 10$^8$ \msun\ \\
    $\dot{M}_{\rm out}$(ionized) & 78$^{+26}_{-25}$ \msunyr\ \\
    $\Sigma_{\dot{M},{\rm out}}$(ionized) & 7~$\pm$~2 \msunyrkpc\ \\
    Ionized fraction & 15~$\pm$~5\% \\ \hline
    \end{tabular}
    \label{table:outflow_props}
\end{table}

Our analysis demonstrates that outflows strong enough to significantly suppress, and potentially even quench, star-formation are present in rapidly assembling massive galaxies in the first billion years of the Universe. With a star formation rate (SFR) of 260~\msunyr, \target\ falls a factor of three above the SFR main sequence and is therefore caught in a phase of relatively fast growth. At the same time, the outflow is removing gas at a rate of \mbox{520 \msunyr}; 20 times faster than typical massive galaxies at this redshift \citep{Ginolfi20a, Birkin25}. Resolved spectral energy distribution (SED) fitting provides tentative evidence that the powerful outflow may have been triggered by a recent burst of star-formation. Clump b, where the outflow likely originates, has a younger stellar age and higher specific SFR than Clumps a, c and d, indicating a larger fraction of young stars (\citealt{Li24}; see Supplementary Material). \target\ may therefore be experiencing a blowout phase triggered by the ongoing galaxy-galaxy interactions. At the very least, this strong outflow is likely to halt the current phase of enhanced star-formation, bringing \target\ back onto the main sequence \citep[e.g.][]{ReichardtChu22b}. If both the star-formation and the outflow continue at their current rates, and the galaxy does not accrete a significant amount of cold gas, the molecular gas reservoir could be exhausted within 50~Myr, leaving a quiescent galaxy with a stellar mass of $\log(M_*/M_\odot)\simeq$~10.5 at $z\sim$~5. 

\target\ has many properties expected for the progenitor of a $z\sim$~3~--~4 massive quiescent galaxy: both observations and simulations suggest that these systems likely formed the bulk of their stellar mass in powerful bursts of star formation at \mbox{$z\sim$~4~--~6}, directly before being rapidly quenched by powerful outflows \citep[e.g.][]{Nanayakkara25, Kimmig25}. A final stellar mass of $\log(M_*/M_\odot)\simeq$~10.5 would place \target\ close to the median for massive quiescent galaxies at $z\sim$~3~--~4 \citep[e.g.][]{Nanayakkara25}. However, \target\ is significantly larger than most massive quiescent galaxies which have typical effective radii $<$~1~kpc \citep[e.g.][]{Carnall24, Ji26}, and it is possible that this system may only quench after the clumps merge. Regardless of the fate of \target, powerful merger-driven outflows may represent a key avenue to produce massive quiescent galaxies at $z\sim$~5. Almost half of massive galaxies at this redshift are experiencing major mergers \citep[e.g.][]{Romano21}. Tidal torques may funnel large amounts of gas towards the galaxy centres, triggering bursts of star-formation and driving powerful outflows which rapidly eject molecular gas, leading to significant suppression of star-formation or even quenching. 

Intriguingly, the powerful outflow from \target\ appears to be driven by stellar feedback, whereas simulations are almost unanimous in their prediction that black hole feedback is needed to create the earliest massive quiescent galaxies \citep[e.g.][]{Kimmig25, Lagos25, Szpila25}. Recent JWST observations have established a link between strong AGN-driven outflows and rapid quenching of massive galaxies at $z\sim$~2~--~3 \citep[e.g.][]{Belli24, Davies24NaD, DEugenio24}. Powerful starburst-driven outflows have been observed in many galaxies spanning the local Universe to high redshifts \citep[e.g.][]{Spilker18, DiamondStanic21, Hagedorn26}. However, determining whether these outflows quench their host galaxies is difficult because starburst-driven outflows are expected to disappear soon after the young stars die \citep[e.g.][]{McQuinn18}, unlike AGN-driven outflows which can persist for several hundred Myr after star-formation ends \citep[e.g.][]{Park24, Bugiani25, Sun26, Taylor26}. Although there is no evidence for current AGN activity at the launch point of the \target\ outflow, we cannot rule out the possibility that the outflow was launched by a powerful AGN that has since faded \citep[e.g.][]{Zubovas23}. 

\target\ illustrates the challenges of robustly identifying and characterising outflows in the early Universe. Our analysis shows that observations probing both rest-frame optical emission lines and neutral gas are required to robustly identify distant outflows and quantify their impact on galaxies. Interacting systems have globally disturbed kinematics, meaning that broad emission line wings in single-slit spectra are not unambiguous signs of outflows \citep[e.g.][]{Parlanti25}. Although \target\ hosts one of the fastest and strongest known star-formation-driven outflows (Fig.~\ref{fig:scaling_relations}), it is only just detectable in the R~$\sim$~1000 JWST spectra (Fig.~\ref{fig:multicomp_fits}). More deep, spatially and spectrally resolved, multiphase observations of massive, high redshift star-forming galaxies are needed to test whether strong, merger-driven outflows are common and whether AGN feedback is required to form early massive quiescent galaxies.

\section*{Acknowledgements}

We thank Sirio Belli and Karl Glazebrook for valuable feedback that improved the clarity of this manuscript. RLD is supported by the Australian Research Council through the Discovery Early Career Researcher Award (DECRA) Fellowship DE240100136 funded by the Australian Government. Parts of this research were supported by the Australian Research Council Centre of Excellence for All Sky Astrophysics in 3 Dimensions (ASTRO 3D), through project number CE170100013. SF acknowledges support from NASA through the NASA Hubble Fellowship grant HST-HF2-51505.001-A awarded by the Space Telescope Science Institute, which is operated by the Association of Universities for Research in Astronomy, Incorporated, under NASA contract NAS5-26555. RA acknowledges financial support from projects PID2023-147386NB-I00 and the Severo Ochoa grant CEX2021-001131-S funded by MCIN/AEI/10.13039/501100011033. MA, RJA, MB, RHC, KT and MS acknowledge support from ANID Basal Project FB210003. MA, RHC and EI acknowledge support from ANID MILENIO NCN2024\_112. MA was supported by FONDECYT grant number 1252054. RJA was supported by FONDECYT grant number 1231718. This work was supported by the French government through the France 2030 investment plan managed by the National Research Agency (ANR), as part of the Initiative of Excellence of Universit\'e Côte d’Azur under reference number ANR-15-IDEX-01. AF acknowledges support from the ERC Advanced Grant INTERSTELLAR H2020/740120. NMFS and H\"U acknowledge funding by the European Union (ERC, GALPHYS, 101055023 and ERC, APEX, 101164796). Views and opinions expressed are, however, those of the author(s) only and do not necessarily reflect those of the European Union or the European Research Council. Neither the European Union nor the granting authority can be held responsible for them. EI acknowledges support from ANID FONDECYT Regular 1221846. RI is supported by Grants-in-Aid for Japan Society for the Promotion of Science (JSPS) Fellows (KAKENHI Grant Number 23KJ1006). GCJ acknowledges support by the Science and Technology Facilities Council (STFC), by the ERC through Advanced Grant 695671 ``QUENCH'', and by the UKRI Frontier Research grant ``RISEandFALL''. IDL acknowledges funding from the European Research Council (ERC) under the European Union's Horizon 2020 research and innovation program DustOrigin (ERC-2019- StG-851622) and from the Flemish Fund for Scientific Research (FWO-Vlaanderen) through the research project G0A1523N. AN and PS acknowledge support from the Narodowe Centrum Nauki (NCN), Poland, through the SONATA BIS grant UMO-2020/38/E/ST9/00077. MR acknowledges support from project PID2023-150178NB-I00 and PID2020-114414GB-I00, financed by MICIU/AEI/10.13039/501100011033, and by FEDER, UE. MS was financially supported by Becas-ANID scholarship \#21221511. K Tadaki acknowledges support from JSPS KAKENHI Grant No. 23K03466. K Telikova acknowledges support from the ALMA-ANID grant 31220026. VV acknowledges support from the ALMA-ANID Postdoctoral Fellowship under the award ASTRO21-0062. WW acknowledges grant support from NASA through JWST-GO-3045 and JWST-GO-3950.

\section*{Data Availability}

The ALMA and JWST observations used in this paper are publicly accessible from online archives. This paper makes use of the following ALMA data: ADS/JAO.ALMA\#2011.0.00064.S, ADS/JAO.ALMA\#2012.1.00523.S, ADS/JAO.ALMA\#2017.1.00428.L, ADS/JAO.ALMA\#2021.1.00280.L. ALMA is a partnership of ESO (representing its member states), NSF (USA) and NINS (Japan), together with NRC (Canada), MOST and ASIAA (Taiwan), and KASI (Republic of Korea), in cooperation with the Republic of Chile. The Joint ALMA Observatory is operated by ESO, AUI/NRAO and NAOJ. This work is based in part on observations made with the NASA/ESA/CSA James Webb Space Telescope. The data were obtained from the Mikulski Archive for Space Telescopes at the Space Telescope Science Institute, which is operated by the Association of Universities for Research in Astronomy, Inc., under NASA contract NAS 5-03127 for JWST. These observations are associated with programs 1727 and 3045. This research is based in part on observations made with the NASA/ESA Hubble Space Telescope obtained from the Space Telescope Science Institute, which is operated by the Association of Universities for Research in Astronomy, Inc., under NASA contract NAS 5–26555. These observations are associated with program(s) 9822 and 13384.


\bibliographystyle{mnras}
\bibliography{mybib} 


\clearpage

\appendix


\input{c2-methods}


\bsp	
\label{lastpage}
\end{document}

%% file: c2-methods.tex
\section*{Supplementary Material}

\renewcommand{\thefigure}{S\arabic{figure}}
\makeatletter
\renewcommand{\theHfigure}{S\arabic{figure}}
\makeatother
\renewcommand{\thetable}{S\arabic{table}}
\makeatletter
\renewcommand{\theHtable}{S\arabic{table}}
\makeatother

\subsection*{ALMA Data}
The ALMA Band 7 products used in this paper combine data from four observing programs. \target\ was observed by program 2011.0.00064.S with a spatial resolution of 0.5'' \citep{Riechers14}, program 2012.1.00523.S ($\theta \sim$~0.6''; \citealt{Capak15}) and the ALPINE survey (2017.1.00428.L; $\theta \sim$~0.67''; \citealt{LeFevre20}). The galaxy was followed-up as part of the CRISTAL survey (\citealt{HerreraCamus25}; 2021.1.00280.L) which observed 19 ALPINE galaxies in both compact (C43-1, $\theta \sim$~0.95'') and extended (C43-4, $\theta \sim$~0.20'') configurations, leading to excellent spatial resolution and a large maximum recoverable scale ($\sim$~4.5'' or $\sim$~28~kpc at $z\sim$~5.3). 

We processed the individual datasets using the appropriate version of \textsc{Casa} and combined the visibilities using the \textsc{concat} task. The continuum was subtracted using the \textsc{uvcontsub} task with \textsc{fitorder=0}. Continuum-subtracted datacubes were produced using Briggs weighting with \mbox{\textsc{robust} = 0.5} and natural weighting, at a \mbox{20~\kms} velocity resolution. The Briggs-weighted cube has a beam full width at half maximum (FWHM) of 0.29~$\times$~0.36'' (1.8 $\times$ 2.2 kpc at $z\sim$~5.3), a spatial sampling of 0.04''/pix and an rms of 160~$\mu$Jy beam$^{-1}$. The natural-weighted cube has a beam FWHM of 0.45~$\times$~0.55'' (2.8 $\times$ 3.4 kpc), a spatial sampling of 0.08''/pix and an rms per 20~\kms\ velocity channel of 130~$\mu$Jy beam$^{-1}$. Cubes were cleaned down to 1$\sigma$ using \textsc{tclean} and moment 0, 1 and 2 maps were produced after applying a blanking mask. To create the mask, the natural-weighted cube was convolved with a \mbox{$\sigma$ = 100~\kms} Gaussian kernel along the velocity axis and a $\sigma$~=~10 pix 2D Gaussian kernel in the spatial axes. The rms was measured from signal-free regions of the convolved cube and a 3D mask was created to isolate pixels above a 2$\sigma$ threshold. This mask was fed into the \textsc{immoments} task to obtain the moment maps from the original cubes.

We robustly detect Band 7 continuum tracing rest-frame $\sim$160$\mu$m emission and use the continuum flux density to compute the molecular gas mass following \citet{Tacconi20}. We assume a dust temperature of 50~K motivated by multi-band FIR SED modelling of CRISTAL-22, a galaxy with similar stellar mass and redshift to \target\ (\mbox{$\log(M_*/M_\odot)$ = 10.4}, $z$~=~5.66; \citealt{Villanueva24}). We adopt a dust emissivity index of $\beta$ = 1.8 and a dust mass-to-light ratio of \mbox{$\alpha_\nu$~=~6.7~$\times$~10$^{19}$ erg s$^{-1}$ Hz$^{-1}$ M$_\odot^{-1}$}, as recommended by \citet{Scoville16}. We estimate the gas-to-dust ratio using the metallicity-dependent prescription from \citet{Genzel15}, where the oxygen abundance is computed from strong emission line ratios as described later. The derived molecular gas mass is $\log(M_{\rm mol}/M_\odot)$~=~10.2, consistent with our estimates based on the narrow component of the \CII\ emission.

\subsection*{JWST/NIRCam and HST Imaging}
\target\ was observed with JWST/NIRCam as part of the COSMOS-Web survey (PID: 1727; \citealt{Casey23}) in the F115W, F150W, F277W, and F444W filters, with an integration time of 1030~s per filter. We reduced the data using the standard JWST Calibration package (version=1.10.0, pmap=1075) and the \textsc{CRAB.Toolkit.JWST} package\footnote{\url{https://github.com/1054/Crab.Toolkit.JWST}}, which includes correction for 1-over-f noise and removal of wisps and snowballs, background removal via the skymatch method, manual masking for varying artifacts like claws, and astrometric alignment to the Gaia DR3 reference with \textit{Tweakreg}.

The \textsc{imfit} code was used to model the JWST F444W image as the convolution of a 2D S\'ersic profile and the point spread function (PSF). The PSF was reconstructed using the \textsc{stpsf} package \citep{Perrin14}. The measured axis ratio was converted to an inclination assuming an intrinsic disk thickness of $q_0$~=~0.25, yielding $i$~=~47$^\circ$.

\begin{figure*}
    \centering
    \includegraphics[scale = 1.1, clip = True, trim = 50 200 130 40]{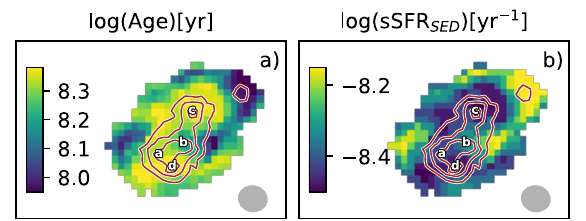} 
    \includegraphics[scale = 1.1, clip = True, trim = 50 200 130 40]{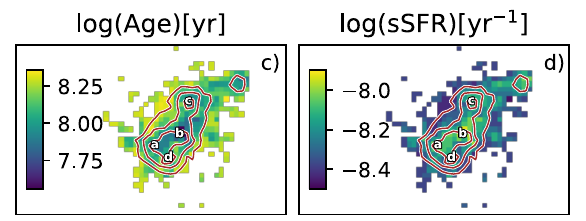} 
    \caption{\textbf{Resolved maps of (left) light-weighted age and (right) specific star-formation rate across \target}, derived from pixel-by-pixel SED fitting. Top panels show results including HST, JWST and ALMA continuum observations, whilst the ALMA data is excluded in the bottom panels. Brown contours show the \Ha\ signal-to-noise ratio at levels of 8, 20 and 40. Both sets of fits suggest that Clump b is younger and has a higher specific SFR than Clumps a, c and d.}
    \label{fig:clump_properties}
\end{figure*}

Spatially resolved SED fitting was performed as described in \citet{Li24}. Existing 5-band HST imaging (F814W, F105W, F125W, F140W and F160W) from programs including COSMOS \citep{Koekemoer07} and CANDELS \citep{Grogin11, Koekemoer11} was reduced and aligned using \textit{grizli} \citep{Brammer23}. Fitting was performed once using the HST and JWST images and a second time including the natural-weighted ALMA Band 7 dust continuum observations. In both cases, all images were re-sampled to the same pixel scale, corrected for misalignments between different filters, homogenised to the same PSF and divided into aperture grids. For all pixels with SNR~$>$~1 in at least 4/10 filters, SED fitting was performed using \textsc{magphys} \citep{DaCunha08}. The star-formation history is parametrized as a continuous delayed exponential function with random bursts to account for stochasticity. Fig. \ref{fig:clump_properties} shows resolved maps of the specific star formation rate (sSFR) and stellar age across \target. Clump b, which falls at the centre of the outflow bicone and is the most likely launching point (see Fig.~\ref{fig:gallery}~ii), has a higher sSFR and lower age than Clumps a, c or d, indicating a larger fraction of young stars. This provides tentative evidence that a recent localised burst of star-formation may have been responsible for triggering the outflow. The integrated SFR and stellar mass estimates obtained by summing the resolved maps are consistent with the global values previously published by \citet{Mitsuhashi24}.

\subsection*{JWST/NIRSpec Integral Field Spectroscopy} 
\target\ was observed with the JWST/NIRSpec Integral Field Unit (IFU) as part of program GO 3045, using the G235M/F170LP and G395M/F290LP grating/filter combinations with integration times of 20 minutes and 82 minutes, respectively. The data were reduced following the standard three-step procedure of the JWST pipeline (version 1.16.0, pmap=1298), utilizing scripts developed by the TEMPLATES team \citep{Rigby25}. During this process, manual masks were applied to mitigate leakage light from intermittently failed open shutters in the MSA, based on visual inspection of all individual exposure frames. As part of the post-processing, we performed additional corrections, including background subtraction using local sky areas within the cubes, empirical removal of systematic stripes, error rescaling, and astrometry alignment. For the astrometry correction, we conducted 2D elliptical Gaussian fitting on the observed NIRCam F277W and F444W maps, as well as pseudo F277W and F444W continuum maps generated by convolving the G235M and G395M cubes with their respective filter response functions. 

The G235M cube covers the \OII~$\lambda \lambda$3726,3729\AA, He~\textsc{ii}~$\lambda$4686\AA, H$\beta$, and \OIII~$\lambda \lambda$4959,5007\AA\ emission lines. The G395M cube also covers the He~\textsc{ii}, H$\beta$ and \OIII\ lines as well as \NII~$\lambda$6548\AA, H$\alpha$, \NII~$\lambda$6583\AA, and \SII~$\lambda \lambda$6716,6731\AA. We use the G395M data in the overlap region due to the four times longer integration time. We perform pixel-by-pixel emission line fitting of the G395M and G235M cubes individually. For each spaxel of each cube, we fit a model consisting of a first order polynomial to account for the stellar continuum and a single Gaussian for each emission line, where all emission lines are assumed to have the same central velocity and intrinsic dispersion. We convolve the intrinsic line profiles with the NIRSpec line spread function (LSF), which we model as a Gaussian with a wavelength-dependent dispersion calculated from the spectral resolution $R$ \mbox{$\left(\sigma_{LSF} (\lambda) = R(\lambda)/2\sqrt{(2 \rm{ln} 2)}\right)$}\footnote{Taken from JDox: \url{https://jwst-docs.stsci.edu/jwst-near-infrared-spectrograph/nirspec-instrumentation/nirspec-dispersers-and-filters}}. All velocity dispersion measurements reported in this paper refer to the intrinsic values after removing instrumental dispersion. We fix the \OIII\ and \NII\ doublet ratios to their theoretical values. 

The \Ha\ flux map is used to compute the \sfrsd\ per pixel. When showing maps derived from emission line measurements, we only include spaxels where the relevant line(s) are detected with \mbox{SNR $>$~5}. The observations are not deep enough to robustly measure \Hb\ fluxes for individual pixels, so we compute the Balmer decrement from the galaxy-integrated spectrum (described in the following section) and correct the observed \Ha\ fluxes for extinction assuming that the starlight is attenuated following the Calzetti curve, the nebular line emission is attenuated following the Cardelli curve, and the unattenuated Balmer decrement is \Ha/\Hb~=~2.86. The measured Balmer decrement is 3.6, corresponding to $A_V$~=~0.43~mag. We convert the pixel \Ha\ luminosities to SFR using the \citet{Kennicutt98} prescription adjusted to the Chabrier IMF, and divide by the pixel projected area to obtain the \sfrsd. 

\begin{figure*}
    \centering
    \includegraphics[scale = 0.8, clip = True, trim = 30 130 0 10]{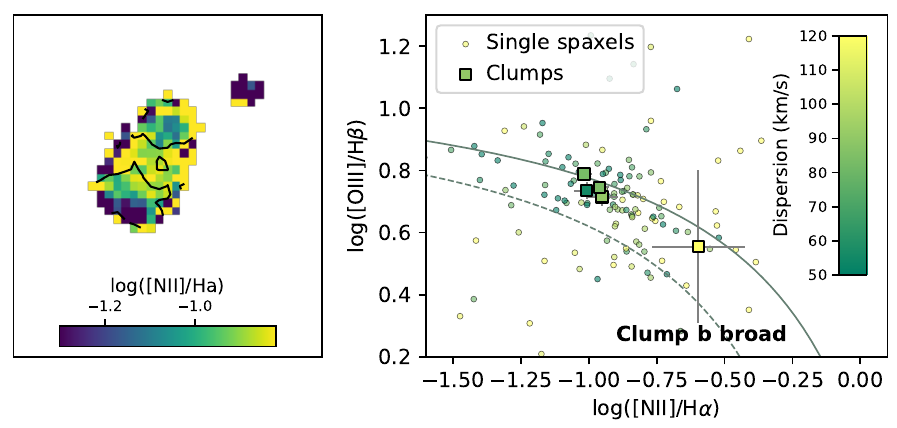} 
    \caption{\textbf{Evidence for shock excitation of the outflowing gas.} Left: \NII/\Ha\ map. Black contours delineate the boundary between the ISM and the outflow cone, defined as \mbox{$\sigma$~=~90~\kms}. The central region of enhanced velocity dispersion shows the largest \NII/\Ha\ ratios. Right: \NII/\Ha\ vs. \OIII/\Hb\ diagnostic diagram for \target. The dashed line is the locus of $z\sim$~2 star-forming galaxies \citep{Strom17} and the solid line shows the maximum line ratios theoretically attributable to pure star-formation \citep{Ke01a}. Small dots represent measurements for individual pixels in which \OIII\ and \Ha\ are detected at SNR~$>$~5 whilst squares with error bars represent the line ratios measured from the aperture spectra of star-forming clumps a, b, c and d. For Clump~b, the line ratios of the narrow ISM component and broad outflow component are plotted separately. The Clump~b outflow component has a significantly higher \NII/\Ha\ ratio than the narrow component and any of the other star-forming clumps. The link between elevated velocity dispersion and enhanced \NII/\Ha\ ratios is indicative of shock excitation.}
    \label{fig:line_ratio_dispersion}
\end{figure*}

\subsection*{Galaxy-integrated spectrum and global properties}
To compute global properties of \target, we extract galaxy-integrated spectra by summing the G235M and G395M cubes over a 2~$\times$~1.6'' elliptical aperture at a position angle of 155 degrees, designed to enclose the main \CII-emitting region (solid white contours in Fig. \ref{fig:gallery}~i). 

We fit the emission lines using a similar procedure as described for the individual pixel spectra. We model the stellar continuum as a first-order polynomial and test two models for the emission lines: a) a single Gaussian component and b) two Gaussian components, with a narrow component tracing the galaxy and a broader component tracing the outflow. When performing the two-component fitting, we found that the best-fit intrinsic velocity dispersion of the broad component (after removing the instrumental broadening) was notably lower when fitting only \NII+\Ha\ compared to when fitting only \OIII+\Hb. Similar discrepancies are seen in other works analysing NIRSpec IFU observations \citep[e.g.][]{Uebler23} and may reflect inaccuracies in the assumed spectral resolution. The intrinsic velocity dispersion measured for the narrow emission line component is similar to the NIRSpec spectral resolution at the observed location of the emission lines ($R\sim$~1000~--~1400 or $\sigma_{\rm LSF} \sim$~90~--~130~\kms). Because the narrow lines are only marginally resolved, the assumed width of the narrow component directly impacts the broad component properties. In addition, the wavelength-dependence of the spectral resolution makes it problematic to tie the velocity dispersions of lines that are widely separated. Therefore, we decided to use separate velocity dispersions for the \NII+\Ha\ and \OIII+\Hb\ complexes in the two-component fitting, whilst still forcing all lines to have the same centroid velocity. We perform the fitting within a Markov Chain Monte Carlo (MCMC) framework using \textsc{emcee} and calculate the Bayesian Information Criterion (BIC; defined as $\chi^2$ + $N_{\rm parameters} \ln(n_{\rm datapoints}$)) to determine whether a broad component is statistically required to explain the observed emission line profile. We find \mbox{BIC(1comp) $-$ BIC(2comp)~$>$ 10} which is canonically interpreted as strong evidence for the presence of a second kinematic component \citep[e.g.][]{ReichardtChu22}. 

We compute the total SFR of \target\ from the extinction-corrected luminosity of the narrow \Ha\ component, using the same method as for the \sfrsd\ map. We obtain a total SFR of 260~$\pm$~30~\msunyr, placing \target\ on the upper envelope of the SFR main sequence \citep{Speagle14}. The \Ha\ SFR is broadly consistent with estimates from SED fitting (\citealt{Mitsuhashi24}: 180$^{+290}_{-110}$~\msunyr; \citealt{Li24} with ALMA: 135$^{+51}_{-28}$~\msunyr, without ALMA: 230$^{+117}_{-98}$~\msunyr). 

We also extract a global \CII\ spectrum, fit two Gaussian components and use the \CII\ luminosity of the narrow component to estimate the molecular gas mass following \citet{Zanella18, Casavecchia25}. We measure a narrow \CII\ luminosity of \mbox{log($L_{\rm [CII],narrow}$/L$_\odot$) = 9.09~$\pm$~0.02}, which corresponds to a molecular gas mass of \mbox{$\log(M_{\rm mol}$/\msun) = 10.6~$\pm$~0.3} using the prescription from \citet{Zanella18} or \mbox{$\log(M_{\rm mol}$/\msun) = 10.3~$\pm$~0.1} using the prescription from \citet{Casavecchia25}. If we use $L_{\rm [CII],narrow}$ to estimate the SFR with the prescription from \citet{Romano22} this yields 110$^{+160}_{-70}$~\msunyr\ (where the errors reflect the intrinsic scatter in the $L_{\rm [CII]}$-SFR relation), also consistent with the \Ha\ measurement.

\subsection*{Clump spectra and properties}
To compute the properties of the outflow and the interstellar medium at the launching site, we extract spectra in a circular aperture centered on Clump b. We choose a diameter of 0.6'' (3.7~kpc) because this is approximately the width of the bicone seen in the ionized gas velocity dispersion map (Fig~\ref{fig:gallery}~ii). However, the outflow velocity and ionized outflow mass do not change significantly when using larger apertures, and the broad component remains even when extracting the spectrum over a single beam. We fit the spectrum of Clump b using the same method as the global spectrum. Once again, the two-component fit is favoured over the one component fit based on the BIC. We note that the best-fit broad components are 2~--~4$\times$ wider than the nominal spectral resolution, and thus uncertainties in the spectral resolution do not have a significant impact on the derived outflow properties. We also extract spectra centered on Clumps~a, c and d using the same aperture size for comparison and perform 1-component fitting to obtain their emission line fluxes and velocity dispersions.

We search for correlations between velocity dispersion and emission line ratios as evidence of shock excitation in the outflowing gas. The left-hand panel of Fig. \ref{fig:line_ratio_dispersion} shows a map of the \NII/\Ha\ ratio for individual spatial pixels across \target. The gas in the outflow cone shows elevated \NII/\Ha\ ratios. The right hand panel shows these pixels in the \NII/\Ha\ vs. \OIII/\Hb\ diagnostic diagram, plotted as small circles color-coded by velocity dispersion. The larger squares show emission line ratios measured from the aperture spectra of clumps a, b, c and d. For Clump~b we separately plot the emission line ratios measured for the narrow ISM component and the broad outflow component. Clumps a, c, d and the narrow component of Clump b have \NII/\Ha\ ratios of $\sim$~0.1, whilst the outflow component in Clump b has a significantly larger \NII/\Ha\ ratio of $\sim$~0.25. The link between elevated velocity dispersion and enhanced \NII/\Ha\ ratios is indicative of shock excitation in the outflowing gas \citep[e.g.][]{Ho14}.

The spectrum of Clump~b does not reveal any evidence for energetically-significant AGN activity. Firstly, the forbidden \OIII\ and \NII\ lines are at least as broad as the permitted Balmer and He~\textsc{i} emission lines, and the velocity of the best-fit broad component is inconsistent with zero at the 4$\sigma$ level (Fig.~\ref{fig:multicomp_fits}). Therefore, replacing the outflow component with an AGN broad line region component results in a notably worse fit. Secondly, the emission line ratios are consistent with star-formation being the primary source of ionization, as shown in Fig. \ref{fig:bpt}. In the \NII/\Ha\ vs. \OIII/\Hb\ diagnostic diagram (left), Clump~b falls in the region occupied by high-redshift star-forming galaxies. However, low metallicity AGN also lie in this region of the diagram, making the classification ambiguous \citep[e.g.][]{Uebler23}. We therefore also make use of the \HeII/\Hb\ diagnostic diagram (right). \HeII\ has an excitation potential of 54 eV and is more sensitive to the presence of a hard ionizing radiation field than \OIII\ which has an excitation potential of 34~eV. We do not detect \HeII\ emission from \target. The upper limit on the \HeII/\Hb\ ratio falls on the upper boundary of the star-forming region, firmly below known AGN host galaxies at similar redshift. \target\ is additionally not detected in VLA-COSMOS radio continuum observations or by Chandra or XMM-Newton in hard X-rays, implying an upper limit on the 2-10~keV X-ray luminosity of $\lesssim$~10$^{44}$~erg~s$^{-1}$.

\begin{figure*}
    \centering
    \includegraphics[scale = 0.7, clip = True, trim = 10 100 10 10]{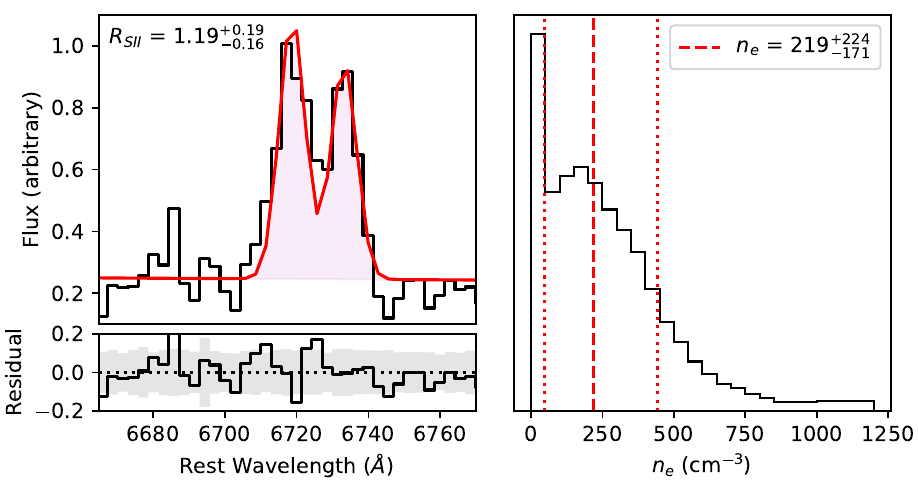} \\
    \caption{\textbf{Measurement of the electron density in \target.} Left: Single component fit to \SII\ emission lines in Clump b. The bottom sub-panel shows the residuals, with the 1$\sigma$ error region shaded in grey. Right: Posterior probability distribution of electron density from MCMC fitting, using the calibration from \citet{Sanders16}. The dashed line indicates the median and dotted lines indicate the 16th and 84th percentile ranges.}
    \label{fig:sii_fit}
\end{figure*}

\begin{sloppypar}
Next, we estimate the gas-phase oxygen abundance at the launching point of the outflow. There is no significant detection of the auroral \OIII~$\lambda$4363\AA\ line, so we use empirical calibrations of strong emission line ratios (\OII/\Hb, \OIII/\Hb, \OIII/\OII, and (\OIII+\OII)/\Hb). We apply the \citet{Sanders24} calibration measured for galaxies at \mbox{2 $<z <$ 9}, yielding \mbox{12 + log(O/H) = 8.35~$\pm$~0.1} \mbox{(0.4~--~0.65~Z$_\odot$)}. This places \target\ on, or slightly above, the mass-metallicity relation at $z\sim$~5.
\end{sloppypar}

Finally, we estimate the electron density at the launching point of the outflow using the \SII~$\lambda \lambda$6716,6731\AA\ doublet ratio which is known to correlate strongly with electron density. The signal-to-noise ratio is not high enough to constrain a meaningful 2-component decomposition, even when fixing the kinematics to the best-fit values measured for the \NII+\Ha\ complex. Therefore, we perform a 1 component fit which primarily probes the ISM density in clump b. The left-hand panel of Fig. \ref{fig:sii_fit} shows the best-fit to the \SII\ doublet profile and the right-hand panel displays the probability distribution for the electron density, constructed by computing the electron density for each value of the \SII\ ratio in the MCMC posterior using  the $n_e$ calibration from \citet{Sanders16}. We measure an \SII\ ratio of 1.19$^{+0.19}_{-0.16}$ which corresponds to an electron density of 219$^{+224}_{-171}$~cm$^{-3}$; similar to electron densities measured for other $z\sim$~5 galaxies with comparable mass and SFR \citep[e.g.][]{Reddy23}.  

\subsection*{Kinematics of the \target\ system}
\begin{figure}
    \centering
    \includegraphics[scale = 0.635, clip = True, trim = 5 160 60 10]{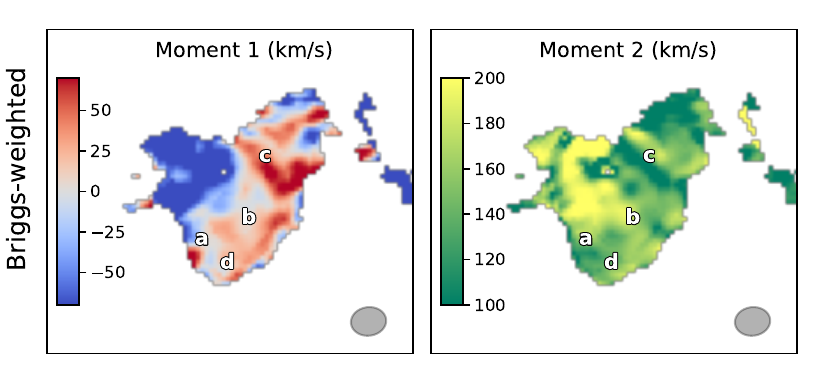} 
    \includegraphics[scale = 0.62, clip = True, trim = 0 160 60 10]{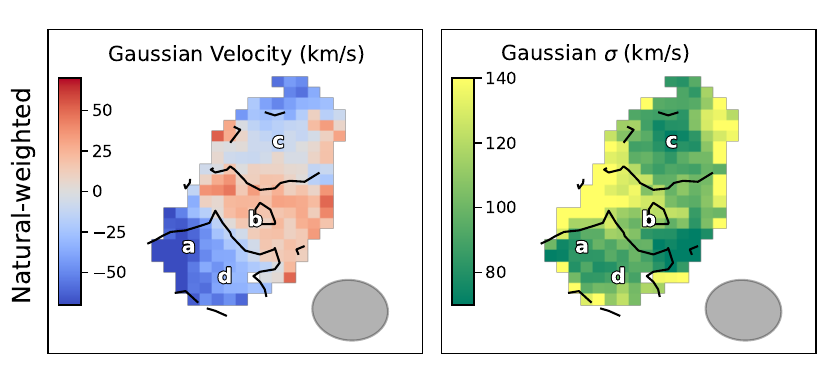} \\
    \caption{\textbf{Kinematics of the \target\ system.} Top: Moment 1 (left) and Moment 2 (right) maps extracted from the Briggs-weighted \CII\ cube. Bottom: velocity (left) and dispersion (right) maps from pixel-by-pixel Gaussian fitting to the Natural-weighted \CII\ cube. The best-fit Gaussian parameters for the plume region are not shown because they have large associated errors due to the weak, broad, blueshifted line profiles in this region. Black contours are the same as in Fig.~\ref{fig:line_ratio_dispersion}. }
    \label{fig:vel_maps}
\end{figure}

To determine the nature of the \target\ system, we examine the resolved kinematics. The top row of Fig. \ref{fig:vel_maps} shows the Moment 1 and Moment 2 maps extracted from the Briggs-weighted \CII\ cube. The gas in the \CII\ plume is clearly blueshifted and shows enhanced dispersion, consistent with expectations for gas on the approaching side of an outflow. The moment maps extracted from Natural-weighted data are not shown but also clearly show the blueshifted, high dispersion plume. There is no obvious rotation pattern across the main body of the galaxy, but the Moment 1 measurements are influenced by the strong asymmetric wings of the \CII\ line profiles (see Fig. \ref{fig:multicomp_fits}). Therefore we also show the velocity map derived from pixel-by-pixel Gaussian fitting of the Natural-weighted \CII\ cube (lower left). There is no data for the plume in the Gaussian \CII\ velocity map because these pixels have relatively low flux spread out over a large velocity range, resulting in large errors on the best-fit Gaussian parameters. As expected, the Gaussian velocity field is significantly more ordered than the Moment 1 map. We see tentative evidence for two velocity gradients, suggesting that \target\ may be comprised of two kinematically distinct components. 

\begin{figure*}
    \centering
    \includegraphics[scale = 0.8, clip = True, trim = 13 80 15 80]{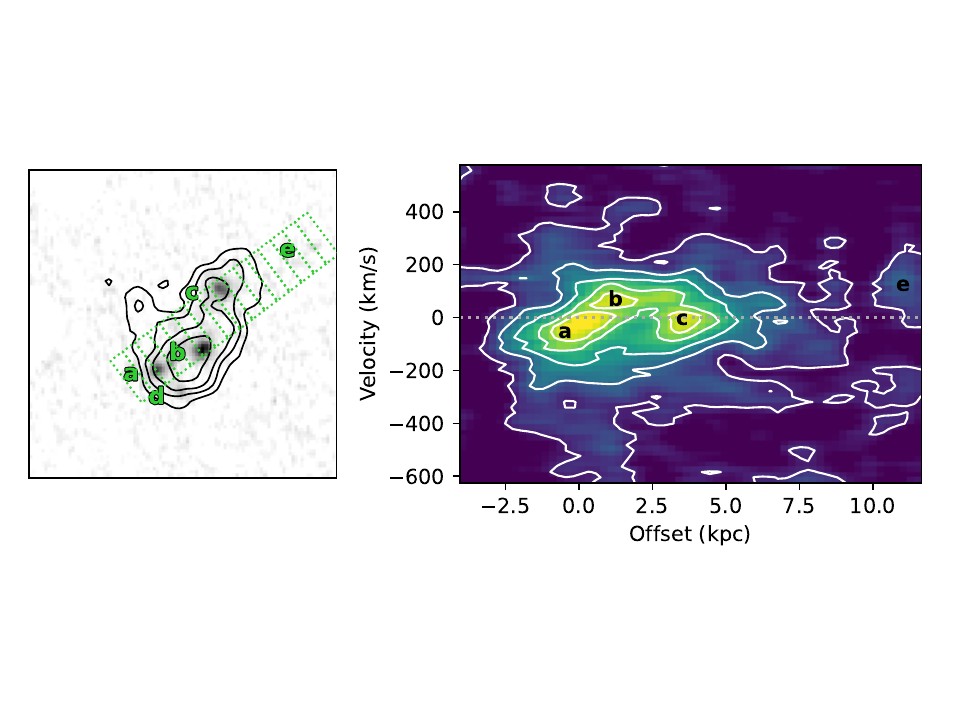}
    \caption{\textbf{Position-velocity diagram extracted along the major axis.} The left panel illustrates the pseudo-slit used to extract the position-velocity diagram, which is oriented at the same angle as the kinematic major axis but shifted upwards to cover Clumps a, b and c. Contours in the left panel are the same as Fig.~\ref{fig:gallery}~i and contours in the right panel are at levels of 1.5, 5, 8, 11 and 13$\sigma$. Clumps a and b are connected in velocity space and could plausibly form a single dynamically-bound structure. Clump c is dynamically distinct and may represent an interacting/merging component. }
    \label{fig:pv_diagrams}
\end{figure*}

To further investigate this possibility, we extract a position-velocity (p-v) diagram. We define the kinematic major axis to be the direction of largest observed velocity difference across the source, which is at a position angle of 127$^\circ$. We construct an 0.7'' wide psuedo-slit with PA~=~127$^\circ$, centered such that it encompasses Clumps~a, b and c (left panel of Fig.~\ref{fig:pv_diagrams}). A p-v diagram is then extracted along this psuedo-slit and smoothed along the velocity axis by averaging over a 5 pixel moving window, as shown in the upper right panel. Clumps a and b are connected by a velocity gradient and could form a single rotating structure, consistent with the disk-like kinematics seen in the Gaussian velocity field. Clumps c and e are kinematically distinct and may represent a separate interacting/merging component. 

To better visualise the spatial distribution of the broad \CII\ emission, we made a channel map summing all \CII\ emission in each pixel within a velocity interval of \mbox{$-$~500~\kms\ $< v < -$~150~\kms}, where the upper boundary represents the velocity below which the outflow dominates the spectrum of Clump~b (Fig.~\ref{fig:multicomp_fits}). The contours in Fig.~\ref{fig:gallery}~ii show that the broad, blueshifted emission is strongly concentrated along the minor axis in the region connecting Clump~b and the plume. Once again, we do not find significant broad, blueshifted emission associated with Clumps a, c or d, providing further evidence that Clump b is the launching point of the outflow. 

\begin{figure*}
    \centering
    \includegraphics[scale = 0.75]{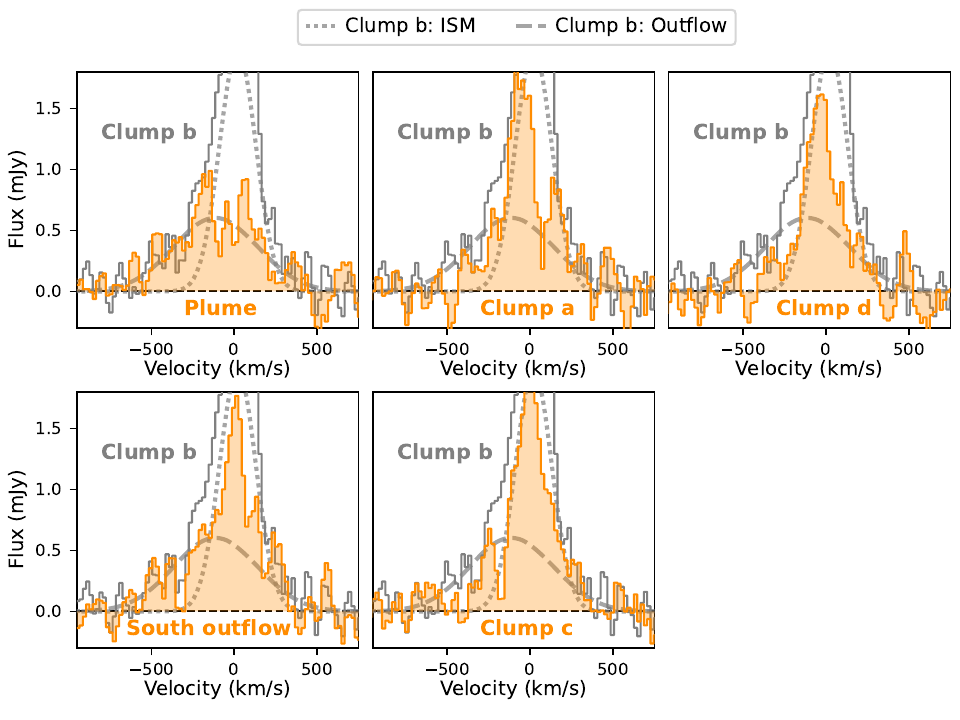} \\
    \caption{\textbf{Spectrum of Clump~b (grey) with the best-fit ISM (dotted) and outflow (dotted) components overlaid, compared to spectra extracted from other regions of \target\ shown as orange shaded regions.} From left to right, top to bottom: the plume, Clump a, Clump d, the south outflow and Clump~c. Strong blueshifted emission is seen in Clump~b, the plume and the south outflow region but is absent in Clumps a, c and d.}
    \label{fig:clump_spectra}
\end{figure*}

To show this more clearly, we examine aperture spectra extracted from the plume region (cyan dashed circle in Fig.~\ref{fig:gallery}~i) as well as Clumps a, c and d. In Fig. \ref{fig:clump_spectra}, each of the 4 panels shows the spectrum of Clump b in grey with the ISM (dotted) and outflow (dashed) components overlaid, compared to spectrum extracted from one of the other apertures (shown by the filled orange regions). The strong broad, blueshifted emission seen in Clump~b is also seen in the plume but is absent in Clumps a, c and d. This provides extra evidence that Clump~b is the launching point of the outflow, and that the outflow is collimated and propagates approximately along the minor axis of the system.

We note that the \CII\ emission from the plume has a very similar shape and amplitude to the broader of the two components fit to the spectrum of Clump~b (Fig.~\ref{fig:multicomp_fits}). This provides additional evidence that the broad emission seen in Clump~b is physically linked to the plume, and also verifies the two-component decomposition of the Clump b spectrum is physically meaningful: the narrow component traces emission from the galaxy (Clump a/c) and the broad component traces the outflow (plume).

We observe consistent dispersion properties between the ionized and \CII-emitting gas (Fig.~\ref{fig:gallery} and Fig.~\ref{fig:vel_maps}, bottom right). The \CII\ dispersion map is qualitatively similar to the ionized gas dispersion map, except for a region of low \CII\ dispersion on the southern side of the outflow cone. To investigate the origin of the low dispersion, we extracted a spectrum of this region, shown in the bottom left panel of Fig.~\ref{fig:clump_spectra}. The southern outflow region shows strong broad emission over a similar velocity range to Clump~b. However, the best-fit single Gaussian velocity dispersion is anomalously low due to the strong narrow peak at $\sim$~0 \kms.

\subsection*{Outflow properties}
\begin{sloppypar}
The properties of the \target\ outflow are listed in Table \ref{table:outflow_props}. We measure a broad \CII\ luminosity of \mbox{L$_{\rm C II,broad}$ = 1.2$^{+0.3}_{-0.2}$~$\times$~10$^{42}$ erg s$^{-1}$}, or \mbox{log(L$_{\rm C II,broad}/L_\odot$)~=~8.5~$\pm$~0.1}. The total mass of \CII-emitting outflowing gas is computed from $_{\rm C II,broad}$ assuming it primarily traces atomic gas in photodissociation regions \citep[e.g.][]{Veilleux20}. The mass-to-light ratio $\kappa$ depends on the gas-phase oxygen abundance, temperature, and the ratio of the gas density to the critical density ($n/n_{\rm crit}$). The gas-phase oxygen abundance of Clump~b is \mbox{12~+~log(O/H) = 8.35} (approximately half solar). Outflows from metal-rich galaxies are expected to have similar metallicity to the ISM \citep[e.g.][]{Chisholm16, Vijayan24}, so we assume the same gas-phase oxygen abundance for the outflow. The mass-to-light ratio is lowest when the density and temperature are the highest, so we conservatively assume \mbox{T~$\simeq$~10$^4$~K} and {$n$~$\simeq$~10$^4$~cm$^{-3}$}. Using the $n_{\rm crit}(T)$ parameterisation for hydrogen atoms from \citet{Goldsmith12}, we obtain \mbox{$\kappa$ = 4.6 M$_\odot$/L$_\odot$} and a total outflow mass of \mbox{$M_{\rm out, [CII]}$ = 1.5~$\pm$~0.3 $\times$ 10$^9$ \msun}. Adopting critical densities representative of ionized or molecular gas changes $\kappa$ by less than 10\%. It is unlikely that the \CII\ luminosity is significantly enhanced by shock excitation: the global log($L_{\rm [CII]}/L_{\rm FIR}$) ratio is -2.49 (using the $L_{\rm FIR}$ reported in \citet{Mitsuhashi24}), consistent with values measured for $z\sim$~0 galaxies with comparable FIR surface brightness \citep{Lutz16} and an order of magnitude lower than values measured for shock-dominated regions \citep{Appleton13}. We compute the mass outflow rate from Clump b assuming a mass-conserving biconical flow, giving \mbox{$\dot{M}_{\rm out}$ = $M_{\rm out} v_{\rm out} / r_{\rm out}$} \citep{Rupke05b}. In this case, $r_{\rm out}$ is the radius of the aperture used to extract the Clump b spectrum (1.9 kpc). Combining all our measurements, we find that the total mass outflow rate from \target\ is \mbox{521$^{+71}_{-89}$~\msunyr}.
\end{sloppypar}

We compute the mass of outflowing ionized gas from the extinction-corrected luminosity of the broad \Ha\ emission following \citet{Davies19}. The Balmer decrement is computed from the sum of the narrow and broad component fluxes because the SNR of the \Hb\ line is too low to robustly constrain a two-component decomposition. We measure a broad \Ha\ luminosity of \mbox{L$_{\rm H\alpha,broad}$ = 7.8$^{+2.8}_{-2.2}$ $\times$~10$^{42}$ erg s$^{-1}$}. The \Ha\ mass-to-light ratio is proportional to $n_e^{-2}$. The ISM electron density at the point of launch of the outflow is 220~cm$^{-3}$ (see Fig. \ref{fig:sii_fit}). Observations and state-of-the-art outflow simulations suggest that the electron density in outflows declines with distance from the midplane \citep[e.g.][]{XuXinFeng23_M82, Schneider20}, although recent observations of edge-on galaxies indicate that the density may increase again at $z \gtrsim$~2~kpc \citep[e.g.][]{MazzilliCiraulo25}. Our Clump~b spectrum probes the central $\sim$~2 kpc of the outflow. We adopt a fiducial electron density of \mbox{$n_e \simeq$~100~cm$^{-3}$}, which is similar to the density of the M82 outflow at \mbox{r $\simeq$ 1 kpc} \citep{XuXinFeng23_M82}. Assuming T~$\simeq$~10$^4$~K and case B recombination, we compute an ionized outflow mass of \mbox{$M_{\rm out, ion} = 2.5^{+0.9}_{-0.7}$ $\times$ 10$^8$ \msun} and an ionized gas outflow rate of \mbox{78$^{+26}_{-25}$~\msunyr}. Ionized gas accounts for 15~$\pm$~5\% of the outflowing mass in \target. If the electron density is 220~cm$^{-3}$ (or even larger as suggested by some previous works; e.g. \citealt{NMFS19}), the ionized outflow mass and the corresponding ionized fraction would decrease by a factor of 2 (or more). However, our main conclusions are based on the \CII\ outflow mass measurements and therefore would not be impacted. In order for ionized gas to contribute $>$~50\% of the outflowing mass, the electron density would have to be very low, \mbox{$n_e$~=~30~cm$^{-3}$}, which is unrealistic given that we are probing the outflow within 1.7~kpc of its point of launch. The outflow mass flux is defined as $\Sigma_{\dot{M}_{\rm out}}\simeq \dot{M}_{\rm out}$/$\pi r_{\rm out}^2$, giving a \CII\ outflow mass flux of \mbox{47$^{+7}_{-8}$ \msunyrkpc} and an ionized outflow mass flux of \mbox{7~$\pm$~2 \msunyrkpc}.

\begin{sloppypar}
Finally, we compute the outflow momentum rate \mbox{($\dot{p}_{\rm out}$ = $\dot{M}_{\rm out} v_{\rm out}$)} and kinetic power (\mbox{$\dot{E}_{\rm out}$ = 0.5 $\dot{M}_{\rm out} v_{\rm out}^2$}), yielding \mbox{$\dot{p}_{\rm out}$ = 2.1~$\pm$~0.4~$\times$~10$^{36}$ dyne} and \mbox{$\dot{E}_{\rm out}$ = 6.5$^{+1.8}_{-1.4}$~$\times$~10$^{43}$ erg s$^{-1}$}. Because we adopt the same velocity for the ionized and neutral phases, the fraction of momentum and energy removed by the ionized phase is the same as the mass fraction (15\%). This is distinct from M82 where the ionized gas is moving $\sim$~2~--~3 times faster than the atomic and molecular phases and therefore removes a significantly larger fraction of energy relative to mass \citep[e.g.][]{XuXinFeng23_M82}.
\end{sloppypar}

We verify that star formation is sufficient to power the observed outflow by comparing the kinetic power and momentum rate of the outflow with the amount of energy and momentum injected into Clump b by supernovae, following \citet{Veilleux20}. From the extinction-corrected luminosity of the narrow \Ha\ emission, we determine that Clump b forms stars at a rate of 177$^{+34}_{-27}$~\msunyr. Assuming a supernova efficiency of 0.01 (i.e. one supernova per 100~\msun\ of stars formed), suitable for a Chabrier IMF, this corresponds to a total momentum injection rate of \mbox{$\dot{p}_{\rm SN}$ = 4.4$^{+0.9}_{-0.7}$~$\times$~10$^{35}$~dyne} and an energy injection rate of \mbox{$\dot{E}_{\rm SN}$ = 6.2$^{+1.2}_{-0.9}$~$\times$~10$^{43}$ erg s$^{-1}$}. The outflow has \mbox{$\dot{E}_{\rm out} \simeq \dot{E}_{\rm SN}$} and \mbox{$\dot{p}_{\rm out} \sim 5 ~\dot{p}_{\rm SN}$}; consistent with values measured for nearby starburst-driven superwinds \citep[e.g.][]{Thompson24}.

We estimate the escape velocity at radius $r$ assuming an isothermal sphere:
\begin{equation}
    v_{\rm esc}(r) \simeq \sqrt{\frac{2 M_{\rm dyn} G \left[1 + \ln(r_{\rm max}/r)\right]}{3r}},
\end{equation}
where M$_{\rm dyn}$ is the dynamical mass and $r_{\rm max}$ is the halo truncation radius. We assume the dynamical mass is 2.5 times the stellar mass \citep{DessaugesZavadsky20}, yielding \mbox{$\log(M_{\rm dyn}/$\msun) $\simeq$~10.7}, and adopt $r_{\rm max}/r$ = 100, consistent with previous works \citep[e.g.][]{Uebler23}. Under these assumptions, the escape velocity at the boundary of the outflow aperture ($r~\simeq$~2~kpc) is \mbox{$\sim$~600~\kms}. Of course, the true mass distribution in the inner regions of \target\ deviates significantly from an isothermal sphere, introducing considerable uncertainty on the escape velocity at small radii. However, the deprojected outflow velocity of 740~\kms\ is considerably higher than the escape velocity. This suggests that a non-negligible fraction of the outflowing material could escape from the \target\ system.

\renewcommand{\arraystretch}{1.2} 
\begin{table*}
    \centering
    \caption{\textbf{Details of the literature comparison samples used in Fig. \ref{fig:scaling_relations} left (global measurements) and right (resolved measurements).} Global measurements are integrated over individual galaxies, except when indicated as stacks. $^a$\citet{Davies19} stacked kpc-scale regions of galaxies in bins of \sfrsd. We show their best-fit power law scaling between \sfrsd\ and $v_{\rm out}$ in the top right panel of Fig. \ref{fig:scaling_relations}. $^b$We use the low-mass stack from \citet{Gupta23} which covers \mbox{log($M_*$/\msun) = 8.4~--~9.0}, but do not include the high-mass stack because the stellar mass range is too large: \mbox{log($M_*$/\msun) = 9.0~--~10.5}. $^c$\citet{NMFS19} and \citet{Concas22} published measurements of star-formation-driven outflows in 4 stellar mass bins. $^d$From \citet{Fluetsch19}, we select all galaxies classified as `HII' with ionized outflow measurements, except M82 which is already included separately. }
    \begin{tabular}{l|l|l|l|l}
    \textbf{Object(s)} & \textbf{Reference(s)} & \textbf{Redshift} & \textbf{Datapoints} & \textbf{Type} \\ \hline
    HZ4	& \citet{HerreraCamus21, Parlanti25} &	5.5 & 1, 1 & resolved, global \\
    GS-4891 & \citet{RodriguezdelPino24} & 3.7 & 1, 1 & resolved, global  \\
    GS-5001	& \citet{Lamperti24} &	3.5 & 1, 1 & resolved, global \\
    zC406690 clumps A \& B & \citet{Newman12_406690} &	2.2  & 2, 1 & resolved, global  \\
    SINFONI-AO resolved stacks$^a$ & \citet{Davies19} & 2.0~--~2.5 & 1 & best-fit line \\
    DUVET & \citet{ReichardtChu25}	& 0.019~--~0.036 & 518, 10 & resolved, global \\
    \multirow{2}{*}{M82}	& \citet{Shopbell98}; &	\multirow{2}{*}{0 (12 Mpc)}  & \multirow{2}{*}{1, 1} & \multirow{2}{*}{resolved, global} \\
    	& \citet{Leroy15, Yuan23} &	  &  &  \\
    \hline
    JADES H$\alpha$ sample	& \citet{Carniani24} &	3.5~--~5.6 & 8 & global \\
    MOSEL EELG stack$^b$ & \citet{Gupta23} &	3~--~4 & 1 & global (stack) \\
    VANDELS & \citet{Llerena23} & 2.2~--~3.5 & 21 & global \\
    MOSDEF & \citet{Weldon24} & 1.4~--~3.8 & 27 & global \\
    KMOS-3D stacks$^c$	& \citet{NMFS19} & 0.6~--~2.7 & 4 & global (stack) \\
    KLEVER stacks$^c$ & \citet{Concas22} &	1.2~--~2.6 & 4 & global (stack) \\
    Makani	& \citet{Rupke19} &	0.46 & 1 & global \\
    Low-$z$ molecular outflows$^d$ & \citet{Fluetsch19} & 0.01~--~0.13  & 6 & global \\
    STARBIRDS Dwarfs & \citet{McQuinn19} &	0 (2~--~5 Mpc) & 11 & global \\ \hline
    \end{tabular}
    \label{table:literature_sample}
\end{table*}

\subsection*{Literature Comparison Samples}
Table \ref{table:literature_sample} summarizes the literature comparison samples used in Fig. \ref{fig:scaling_relations}. The left-hand panel of Fig. \ref{fig:scaling_relations} shows ionized outflow mass loading factor as a function of galaxy stellar mass, and we include measurements for individual galaxies as well as measurements from stacked spectra of galaxies binned by stellar mass. The right-hand panels of Fig. \ref{fig:scaling_relations} show outflow velocity and ionized mass flux as a function of \sfrsd\ and therefore only includes galaxies for which spatially resolved outflow measurements are available. The main goal of these panels is to examine whether the properties of the \target\ outflow are consistent with the extrapolation of the $z\sim$~0 trends to higher \sfrsd. Therefore, we re-calculate the ionized mass flux using the assumptions adopted by \citet{ReichardtChu22} for the $z\sim$~0 measurements. Specifically, we change the assumed outflow radius from 1.9 kpc (the size of the extraction aperture) to 500 pc, thus increasing the outflow mass flux by a factor of 3.5. In the bottom right panel of Fig.~\ref{fig:scaling_relations}, the open star shows the fiducial measurement and the filled star shows the homogenized measurement.